\documentstyle[12pt]{article}

\def\dfrac#1#2{{\displaystyle {#1 \over #2}}}
\def\dsum{\mathop{\displaystyle \sum }}
\def\dint{\displaystyle \int }

\renewcommand{\arraystretch}{1.5} 

\begin{document}

\begin{titlepage}
\noindent BU HEP-96-19 \hfill July 1996\\

\begin{center}

\vspace{0.2cm}

{\Large\bf QCD chiral Lagrangian on the lattice, strong \\
\vspace{0.2cm} coupling expansion and Ward identities with \\
\vspace{0.4cm} Wilson fermions} 

\vspace{0.8cm}

{\bf A.~R.~Levi, V.~Lubicz, C.~Rebbi }

\vspace{0.8cm}

{\it Department of Physics, Boston University, \\
{\rm 590} Commonwealth Avenue, Boston, MA {\rm 02215}, USA }

\end{center}

\vspace{1.5cm}

\abstract{ \par\noindent
We discuss a general strategy to compute the coefficients of the QCD
chiral Lagrangian using lattice QCD with Wilson fermions. This procedure
requires the introduction of a lattice chiral Lagrangian as an
intermediate step in the calculation. The QCD chiral Lagrangian is then
obtained by expanding the lattice effective theory in increasing powers
of the lattice spacing and the external momenta. In order to investigate
the consequences of the chiral symmetry breaking induced by the Wilson
term, we study the lattice chiral Lagrangian at the leading order of the
strong coupling and large $N$ expansion. We show that the effects of the
Wilson term can be conveniently taken into account, in the lattice
effective theory, by a suitable renormalization procedure. In
particular, we show that, at the leading order of the strong coupling
and large $N$ expansion, the chiral symmetry is exactly recovered on the
lattice provided that the bare quark mass and the lattice operators are
properly renormalized.} 

\vspace{1.5cm}

\vfill
\end{titlepage}

\baselineskip 18pt plus 2pt minus 2pt


\section{Introduction\label{sec:intro}}

The QCD Chiral Lagrangian, originally introduced by Weinberg
\cite{wein1} as a convenient way to reproduce the predictions of PCAC
and current algebra, has been subsequently promoted to the role of a
consistently renormalizable effective theory \cite{wein2,gl}, which
provides a powerful tool for describing the phenomenology of low-energy
QCD. A recent review of the phenomenological results obtained in this
framework can be found in refs.~\cite{meiss}-\cite{ecker}.

The basic observation of the chiral Lagrangian approach is that, at very
low energy, the light pseudoscalar mesons are the only hadronic
particles which can be close to the mass shell in the strong interaction
processes, and can therefore contribute poles to the analytic structure
of the amplitudes. Once these contributions are accounted for by the
introduction of a suitable set of fields, an expansion in powers of
momenta becomes possible. The QCD chiral Lagrangian is then the most
general effective theory, constructed in terms of pseudoscalar fields
and a set of external sources, which is consistent with QCD chiral
symmetry and its assumed spontaneous dynamical breaking.

Another important feature of low-energy strong interactions is that, in
the limit of massless light quarks ($u$, $d$, $s$), chiral symmetry
prevents the interactions among the pseudoscalar Goldstone bosons at
zero external momenta. Therefore, the effective theory can be
systematically organized as an expansion in increasing powers of the
quark masses and external momenta: 
\begin{equation}
\label{l2l4}
{\cal L}_{eff} = {\cal L}_2 + {\cal L}_4 + \ldots
\end{equation}

The most general form of the chiral Lagrangian is constrained by the
symmetries of the fundamental QCD theory. This fixes completely the form
of the interactions among the pseudo-Goldstone bosons and the
correlation functions of the external currents in the low-energy
limit. However, a set of numerical coefficients, which define the
strength of these couplings, cannot be fixed by symmetry requirements
alone. 

At the lowest order $p^2$, the effective Lagrangian ${\cal L}_2$ can be
expressed in terms of only two arbitrary parameters which represent, in
the chiral limit, the pseudoscalar decay constant $F_{\pi}$ and the
quark condensate. One finds then that ten additional chiral couplings
are necessary to describe the low-energy QCD phenomenology at ${\cal O}
(p^4)$ \cite{gl}. They are usually denoted by the symbols $L_1, \ldots
, L_{10}$. The main purpose of this paper is to discuss a possible first 
principle lattice calculation of these coefficients.

At present, our knowledge of the coefficients of QCD chiral Lagrangian
mainly comes from the experimental data. The couplings are fixed by
comparing the predictions of the effective theory with a set of physical
amplitudes measured in the experiments. Further constraints on the
coefficients can be derived by general theoretical considerations,
relying, for instance, on the large $N$ expansion. Such a determination
is typically affected by large uncertainties, and some of these
coefficients are still known with a relative error larger than $100\%$
\cite{beg}.

From a theoretical point of view, the coefficients of chiral Lagrangian
ought to be calculable from the QCD Lagrangian. They are functions of
the fundamental QCD scale, $\Lambda_{QCD}$, and the heavy quark
masses. A rough estimate of their size can be obtained by requiring that
the tree-level amplitudes from ${\cal L}_4$ are of the same order of
magnitude of the loop corrections from ${\cal L}_2$ \cite{gm}. This
leads to the prediction $L_i \sim (F_{\pi}^2/4)/ \Lambda_{\chi}^2 \sim 2
\times 10^{-3}$, where $\Lambda_{\chi} \simeq 4 \pi F_{\pi}$ is the
scale of chiral symmetry breaking. Such an estimate reproduce in fact
the correct order of magnitude of the chiral coefficients renormalized
at the rho mass scale, indicating a good convergence of the momentum
expansion below the resonance region.

A more quantitative picture can be obtained by exploiting the role of
meson resonances in the low-energy effective theory \cite{res}. The
basic idea consists in writing down the most general low-energy
effective Lagrangian for effective vector, axial, scalar, pseudoscalar
and tensor fields. The standard chiral Lagrangian is then obtained by
integrating out the mesonic degrees of freedom but for the pseudoscalar
ones. In this way, one ends up with an explicit expression for the
chiral coefficients in terms of low-energy mesonic parameters, masses
and decay constants. This approach provides values of the chiral
couplings in good agreement with the experimental determinations,
showing that these couplings are almost completely saturated by the
mesonic resonances contributions. However, in order to determine their
actual values, one must still rely on the experimental measurements. 

In principle, the QCD chiral Lagrangian can be obtained by integrating
out the non effective quark and gluonic degrees of freedom from the
original fundamental theory. Indeed, being the two theories
mathematically equivalent, the partition function $Z$ can be expressed
as: 
\begin{equation}
\label{zz}
Z=\int \left( dA d\psi d\overline{\psi }\right) \, \exp \left\{ i \int
d^4 x \, {\cal L}_{QCD} \right\} = \int \left( dU \right) \, \exp \left\{ i 
\int d^4 x \, {\cal L}_{eff} \right\}
\end{equation}
However, although the integral over the fermionic fields in the above
equation can be performed explicitly, we do not know, in the general
case, how to perform analytically the remaining integration over the
gluonic fields. 

A more feasible theoretical approach would consist in performing a
matching between the effective and the fundamental theory. Specifically,
one could compute a sufficient number of physical amplitudes, both in
the effective and the original theory, and derive the values of the
chiral coefficients from a comparison of the results. The calculation in
the full theory, being non perturbative, would then require the
implementation of a numerical lattice simulation. 

However, such an approach would be affected by the following technical
difficulty: in order to perform the matching between the effective and
the full theory, one should consider a set physical amplitudes defined
in the region of low external momenta, $p < m_{\rho}$, where the
predictions of the chiral Lagrangian can be reliably obtained. On the
other hand, in the numerical calculation, an intrinsic infra-red cut-off
is introduced by the finite size of the lattice. The minimum value of
momentum that can be considered is $p_{min}=2\pi/La$, where $L$ is the
lattice size and $a$ the lattice spacing. In current lattice
calculations, $p_{min}$ is typically of the order of the rho mass and
thus lies at the border of the kinematical region accessible to the
calculations in the effective theory. In order to overcome this
difficulty, one should consider either larger values of the lattice 
spacing, thus increasing the finite cut-off effects on the lattice, or
larger lattices, which soon becomes computationally prohibitive.

An alternative approach to the calculation of the coefficients of the
chiral Lagrangian has been suggested in ref.~\cite{cs}. The basic
observation there is that the separation between effective and
non-effective degrees of freedom, which occurs in the continuum QCD
theory, must be mirrored by an equivalent distinction in theory
regularized on the lattice. Specifically, one can consider an effective
Lagrangian, defined on the lattice in terms of an effective pseudoscalar
field and external sources, which is equivalent to the fundamental QCD
theory regularized on the lattice, for any value of the lattice spacing
or bare coupling constant.

This effective lattice theory can be then introduced as an intermediate
step in the calculation of the continuum chiral Lagrangian. In order to
derive the effective lattice Lagrangian, one can assume a sufficiently
large set of couplings and fix the corresponding coefficients through a
matching with an overcomplete set of expectation values, computed both
in the effective and the fundamental lattice theory. The several
interactions, allowed in the lattice effective Lagrangian, can be
organized as an expansion in terms of the distance of couplings, rather
than in powers of the external momenta as in the continuum effective
theory. Therefore, the lattice effective theory is not specifically a
low-energy effective theory. This means that, in performing the matching
between the two lattice theories, the existence of an infrared cut-off
in the numerical simulation should not represent a problem any longer. 

A second advantage of this procedure is that the matching is performed
between theories which are defined in the same regularization scheme,
the four-dimensional lattice grid. For this reason, the finite
ultra-violet cut-off effects, which affect the results of the numerical
simulation performed with the fundamental lattice theory, can be kept
better under control, because these effects are in principle exactly
predicted and reproduced by the corresponding lattice effective theory. 

An important observation of ref.~\cite{cs} is that the lattice effective
theory explicitly contains the collective fields which are responsible
for the long distance behavior of the corresponding fundamental
theory. For this reason, only short distance couplings, typically on the
scale length of the order of the inverse rho mass, are expected to play
a significant role in the effective theory. This observation then
provides a criterion to select a finite number of possible interactions
to be considered in the lattice effective theory. For the same reason,
one expects that the determination of the corresponding coefficients
should be feasible on a lattice of moderate size, thus allowing to
achieve a better numerical accuracy in the calculation. 

Once the matching has been performed, and the lattice effective theory
has been derived, one can consider the infinite volume limit of this
theory and expand in increasing powers of the external momenta. The
result of this expansion is the QCD chiral Lagrangian. In this way, the
chiral coefficients can be calculated.

A useful feature in this approach is that, in the limit of strong
coupling and large number of colors $N$, the integration (\ref{zz}) of
the non effective degrees of freedom in the QCD Lagrangian can be
performed analytically on the lattice \cite{klu,ks}. In this way, at the
leading order of the strong coupling and large $N$ expansion, the
lattice effective theory can be exactly computed. 

This procedure has been followed by the authors of ref.~\cite{cs} in
order to compute the coefficients of the QCD chiral Lagrangian in the
strong coupling and large $N$ limit. They have considered a body
centered hypercubical lattice, whose greater symmetry implies that
invariance under lattice transformations carries over to Euclidean
invariance up to the terms of ${\cal O}(p^4)$ in the continuum. Despite
the strong coupling and large $N$ approximation, and an additional
simplifying assumption of complete decoupling of the mesonic resonances,
the final results for the chiral coefficients are in remarkable
qualitative agreement with the experimental values. These results
strongly encourage the attempt of a numerical calculation performed in
the region of intermediate couplings, which is relevant for continuum
QCD calculations. 

The feasibility of such a calculation is further investigated in this
paper. With respect to ref.~\cite{cs}, we consider the calculation on a
standard hypercubical lattice, whose symmetry properties are well
understood and four-dimensional Euclidean symmetry is known to be
properly recovered in the continuum limit. Moreover, in this paper we
study a fermionic action with the Wilson term \cite{wilson}, which
serves to prevent the appearance of the undesired doubler fermionic
species. A well known effect of the Wilson term on the lattice is the
introduction of an additional source of chiral symmetry breaking, which
persists even in the limit of vanishing bare quark masses. Main goal of
this paper is a study of the effects of such a symmetry breaking from
the specific point of view of the calculation of the coefficients of
chiral Lagrangian, along the lines discussed above. 

In order to derive the low-energy effective theory for QCD, it is
convenient to introduce a given set of external source in the
fundamental Lagrangian \cite{gl}. In this way, the original global
chiral symmetry of QCD can be formally promoted to an exact local
invariance. On the lattice, because of the presence of the Wilson term
in the fermionic action, it is necessary to introduce an additional set
of external sources, which have no direct continuum correspondence. This
step is discussed in sec.~\ref{sec:sources}, where we define all the
external sources added to the action and the corresponding lattice
operators which these sources are coupled with. 

In sec.~\ref{sec:eff-action}, we derive the corresponding lattice
effective Lagrangian, at the leading order in the strong coupling and
large $N$ expansion, by following the procedure of
refs. \cite{cs}-\cite{ks}. The main purpose of this calculation is to
investigate the general structure of the lattice effective theory, in
the presence of the Wilson term and the external sources. This general
structure is a pure consequence of chiral invariance and discrete
space-time symmetries, and must persist even in the region of
intermediate lattice couplings. Therefore, the lattice effective
Lagrangian, as derived in the strong coupling limit, can be assumed as a
guideline to define the corresponding theory in the region of
intermediate couplings.

A primary effect of the presence of the Wilson term in the fermionic
action is the appearance, in the lattice effective theory, of additional
terms, which do not have any direct correspondence in the continuum
chiral Lagrangian. The role of these terms is mainly to reproduce the
effects of chiral symmetry breaking induced by the Wilson term. These
effects are first investigated in sec.~\ref{sec:CF}, where some typical
physical quantities, like the pseudoscalar mass, the decay constant and
the kaon $B_K$-parameter, are computed directly from the strong coupling
lattice effective Lagrangian. 

The effects of chiral symmetry breaking are then further investigated in
sec.~\ref{sec:ward}, where we study the axial and vector Ward identities
in the lattice theory. For the weak coupling region, these identities
have been studied in ref.~\cite{bochicchio}. In that paper it was shown
that, once the chiral limit of the theory is correctly identified and
the lattice operators are properly renormalized, the predictions of
continuum PCAC and current algebra are reproduced on the lattice up to
vanishing cut-off effects, unaffected by the chiral breaking introduced
by regularization. In this paper, we show that the same conclusion also
applies to the lattice theory at the leading order of the strong
coupling and large $N$ expansion. In this limit, the effects of the
Wilson terms in the effective Lagrangian can be taken into account
through a proper renormalization of the quark mass and the lattice
operators. The correlation functions of these renormalized operators
then satisfy, even in the strong coupling limit, the Ward identities of
the continuum theory, as predicted by recovered chiral symmetry.

In the effective continuum limit, that is in the limit of vanishing
lattice bare coupling constant, the effects of chiral symmetry breaking
induced by the Wilson term are expected to become completely
negligible. However, it is well known that for typical values of
couplings considered in current numerical simulations these effects are
still quite relevant. For this reason, the role of the ``Wilson terms'',
in the lattice effective Lagrangian, cannot be neglected. On the other
hand, the presence of these terms introduces additional
difficulties. They significantly increase the number of couplings in the
effective theory, they do not have a direct continuum limit in the QCD
chiral Lagrangian and finally, because of their presence, the external
sources in the lattice theory do not reduce straightforwardly to their
direct continuum counterparts. 

A procedure which allows to completely discard the Wilson terms in the
lattice effective theory is then described in sec.~\ref{sec:rino}. In
this way, the general strategy discussed in this paper to derive the QCD
chiral Lagrangian should be significantly simplified. The basic idea is
to consider a new ``renormalized'' effective lattice Lagrangian, which
reproduce the correlation functions of properly renormalized lattice
operators in the fundamental theory. Because these correlation functions
satisfy all the Ward identities predicted by continuum current algebra,
the renormalized effective Lagrangian does not contain the Wilson terms
at all. The way in which this mechanism works is illustrated through the
discussion of few significative examples in the framework of the strong
coupling effective theory, which exhibits the same chiral structure
expected in the weak coupling limit. 

Finally, in sec.~\ref{sec:scaling} we briefly discuss some aspects of
the numerical calculation to be performed in the region of intermediate
couplings of lattice QCD and present our conclusions.

\section{The chiral invariant Wilson action with external sources\label
{sec:sources}}

According to Gasser and Leutwyler \cite{gl}, a convenient preliminary
step in the derivation of the QCD chiral Lagrangian consists in adding
to the original QCD action a proper set of external sources. These
sources are introduced in such a way that the resulting action becomes
invariant with respect to combined local chiral transformations of the
external sources and fundamental fields. In this section, we perform
this step for QCD regularized on the lattice. But for the introduction
of the Wilson term in the lattice action, our definitions of the
external sources closely follow those of ref.~\cite{cs}.

We consider the Wilson formulation of lattice QCD \cite{wilson}. The
total action is: 
\begin{equation}
\label{esse} S=S_g+S_\psi 
\end{equation}
where $S_g$ is the pure gauge action: 
\begin{equation}
\label{Sg} S_g=-\frac 1 {g^2}\sum_P{\rm Tr}\left(
U_P+U_P^{\dagger}\right) 
\end{equation}
$U_P$ being the plaquette variable and the sum over $P$ running over all
the oriented plaquettes on the lattice, and $S_\psi $ is the fermionic
action:
\begin{equation}
\label{Spsi} 
\begin{array}{l}
S_\psi =-\dfrac 12\dsum\limits_{x,\mu }\left[ 
\overline{\psi }(x)\left( r-\gamma _\mu \right) U_\mu (x)\psi (x+\widehat{%
\mu })+\overline{\psi }(x+\widehat{\mu })\left( r+\gamma _\mu \right) U_\mu
^{\dagger }(x)\psi (x)\right] + \\ \qquad +\ \left( m+4r\right)
\dsum\limits_x^{}\,\left[ \overline{\psi }(x)\psi (x)\right] \, 
\end{array}
\end{equation}
The quark fields, $\psi $ and $\overline{\psi }$, are $n-$component
vectors in flavour space; $U_\mu $ are the lattice gauge variables,
defined on the links, and $m$ is the bare quark mass. For simplicity, we
assume in this paper that the $n$ flavours of quarks are degenerate in
mass. 

In the limit of zero quark mass and vanishing Wilson term $(m=r=0)$, the
lattice action is invariant with respect to global chiral
transformations of the quark fields: $\psi _L(x)\rightarrow L\,\psi
_L(x)$ and $\psi _R(x)\rightarrow R\,\psi _R(x)$, with $L$ and $R$
constant unitary matrices in flavour space. Formally, by adding to the
fermionic action a proper set of external sources, this global
invariance can be promoted to a local invariance, even in the presence
of quark masses and Wilson term. We consider a pair of ``link-type''
external sources, $W_\mu (x)$ and $Z_\mu (x)$, defined on the link
between $x$ and $x+\widehat{\mu }$, and a pair of local sources, $\xi
(x)$ and $\chi (x)$, defined on the site $x$. The sources $W_\mu (x)$
and $Z_\mu (x)$ are introduced to perform the parallel transport of the
right- and left-handed quark fields respectively. The sources $\xi (x)$
and $\chi (x)$, which include the Wilson and mass term respectively, and
are introduced to achieve chiral invariance of these terms as well. The
lattice fermionic action, including the external sources, can be written
as:
\begin{equation}
\label{Spsij}
\begin{array}{l}
S_{\psi j}=-\dfrac 12\dsum\limits_{x,\mu }\left\{ 
\overline{\psi }(x)\left[ rK(x)-\gamma _\mu \right] U_\mu (x)Y_\mu (x)\psi
(x+\widehat{\mu })\right. \ + \\ \qquad +\left. 
\overline{\psi }(x+\widehat{\mu })\left[ rK(x+\widehat{\mu })+\gamma _\mu
\right] U_\mu ^{\dagger }(x)Y_\mu ^{\dagger }(x)\psi (x)\,\right\} \ + \\ 
\qquad +\ \dsum\limits_x^{}\left[ \,\overline{\psi }(x)J(x)\psi (x)\,\right] 
\end{array}
\end{equation}
where the external sources, $Y_\mu(x)$, $K(x)$ and $J(x)$ are given by: 
\renewcommand{\arraystretch}{2.0} 
\begin{equation}
\label{sources}
\begin{array}{cll}
Y_\mu (x) & = & \dfrac 12\left[ W_\mu (x)\,(1+\gamma _5)+Z_\mu
(x)\,(1-\gamma _5)\right]  \\ 
K(x) & = & \dfrac 12\left[ \xi ^{\dagger }(x)\,(1+\gamma _5)+\xi
(x)\,(1-\gamma _5)\right]  \\ 
J(x) & = & \dfrac 12\left[ \chi ^{\dagger }(x)\,(1+\gamma _5)+\chi
(x)\,(1-\gamma _5)\right] 
\end{array}
\end{equation}
\renewcommand{\arraystretch}{1.5}$W_\mu(x)$, $Z_\mu(x)$, $\xi(x)$ and 
$\chi(x)$ are matrices in the pure flavour space.  

It is easy to verify that the total action, $S_g+S_{\psi j}$, is now 
invariant with respect to local chiral transformations of the quark
fields:
\begin{equation}
\label{psi-t}\psi _L(x)\rightarrow L(x)\,\psi _L(x)\qquad ,\qquad \psi
_R(x)\rightarrow R(x)\,\psi _R(x)
\end{equation}
with $L(x)$ and $R(x)$ unitary matrices, provided the external sources
are defined to transform according to: 
\begin{equation}
\label{sources-t}
\begin{array}{ccl}
W_\mu (x) & \rightarrow  & R(x)W_\mu (x)\,R^{\dagger }(x+
\widehat{\mu }) \\ Z_\mu (x) & \rightarrow  & L(x)Z_\mu (x)\,L^{\dagger }(x+
\widehat{\mu }) \\ \xi (x) & \rightarrow  & R(x)\xi (x)\,L^{\dagger }(x) \\ 
\chi (x) & \rightarrow  & R(x)\chi (x)\,L^{\dagger }(x)
\end{array}
\end{equation}

The Wilson standard fermionic action (\ref{Spsi}) is recovered from
$S_{\psi j}$ in eq.~(\ref{Spsij}) by considering the limit: 
\begin{equation}
\label{limit}
\begin{array}{c}
W_\mu (x),\ Z_\mu (x),\ \xi (x),\ \xi ^{\dagger }(x)\ \rightarrow \ 1 \\ 
\chi (x),\ \chi ^{\dagger }(x)\ \rightarrow \ m+4r
\end{array}
\end{equation}
In the following, we will refer to this limit as the limit of vanishing
external sources. This limit will be also indicated with the symbol
$\left( \ldots \right) _0$.

By following ref.~\cite{cs}, we express, in a non linear way, the
external fields $W_\mu (x)$ and $Z_\mu (x)$ in terms of traceless
hermitian vector and axial sources: 
\begin{equation}
\label{va-sources}
\begin{array}{c}
W_\mu (x)=\exp \left\{ -i\left[ v_\mu (x)+a_\mu (x)\right] \right\}  \\ 
Z_\mu (x)=\exp \left\{ -i\left[ v_\mu (x)-a_\mu (x)\right] \right\} 
\end{array}
\end{equation}
where: 
\begin{equation}
\label{va-0k}v_\mu (x)=v_\mu ^k(x)\ t^k\qquad ,\qquad a_\mu (x)=a_\mu ^k(x)\
t^k
\end{equation}
and $t^k$ are the generators of the $SU(n)$ flavour group ($k=1,\ldots
,n^2-1$). In the fermionic action (\ref{Spsij}), the sources $v_\mu $
and $a_\mu $ are directly coupled to non-singlet vector and axial
lattice currents. In the limit of vanishing external sources, the
functional derivatives of the action with respect to $v_\mu $ and $a_\mu
$ are given by:  
\begin{equation}
\label{va-def}\left( \dfrac{i\,\delta S}{\delta v_\mu ^k(x)}\right) _0=V_\mu
^k(x)\qquad ,\qquad \left( \dfrac{i\,\delta S}{\delta a_\mu ^k(x)}\right)
_0=A_\mu ^k(x)
\end{equation}
where 
\begin{equation}
\label{v-current}
\begin{array}{l}
V_\mu ^k(x)=\dfrac 12\left[ 
\overline{\psi }(x)U_\mu (x)\left( \gamma _\mu -r\right) t^k\,\psi (x+
\widehat{\mu })\ +\right.  \\ \qquad \qquad \left. +\ \overline{\psi }(x+
\widehat{\mu })U_\mu ^{\dagger }(x)\left( \gamma _\mu +r\right) t^k\,\psi
(x)\right] 
\end{array}
\end{equation}
and 
\begin{equation}
\label{a-current}
\begin{array}{l}
A_\mu ^k(x)=\dfrac 12\left[ 
\overline{\psi }(x)U_\mu (x)\left( \gamma _\mu -r\right) \gamma _5t^k\,\psi
(x+\widehat{\mu })\ +\ \right.  \\ \qquad \qquad \left. +\ \overline{\psi }%
(x+\widehat{\mu })U_\mu ^{\dagger }(x)\left( \gamma _\mu +r\right) \gamma
_5t^k\,\psi (x)\right] 
\end{array}
\end{equation}
$V_\mu ^k(x)$ is the extended lattice vector current, conserved for
degenerate quark masses, and $A_\mu ^k(x)$ is the corresponding (non
conserved) axial current.

Similarly, we decompose the external fields $\xi (x)$ and $\chi (x)$ in 
terms of hermitian scalar and pseudoscalar sources: 
\begin{equation}
\label{sp-sources}
\xi (x)=\sigma (x)+i\pi(x) \qquad ,\qquad \chi (x)=s(x)+ip(x)
\end{equation}
Derivatives of the action with respect to these sources are given by: 
\renewcommand{\arraystretch}{2.5} 
\begin{equation}
\label{sp-def}
\begin{array}{lcl}
\left( \dfrac{\delta S}{\delta s^\alpha (x)}\right) _0=S^\alpha (x) & , & 
\quad \left( 
\dfrac{i\,\delta S}{\delta p^\alpha (x)}\right) _0=P^\alpha (x) \\ \left( 
\dfrac{\delta S}{\delta \sigma ^\alpha (x)}\right) _0=-4\,r\,\Sigma ^\alpha
(x)\quad  & , & \quad \left( \dfrac{i\,\delta S}{\delta \pi ^\alpha (x)}%
\right) _0=-4\,r\,\Pi ^\alpha (x)
\end{array}
\end{equation}
\renewcommand{\arraystretch}{1.5}where $S(x)$ and $P(x)$ are the local
scalar and pseudoscalar densities: 
\begin{equation}
\label{sp1-curr}
S^\alpha (x)=\overline{\psi }(x)t^\alpha \psi (x)\qquad ,\qquad 
P^\alpha (x)=\overline{\psi }(x)t^\alpha \gamma _5\psi (x)
\end{equation}
($\alpha =0,1,\ldots ,n^2-1$ and we use the notation $t^0=1$). The
operators $\Sigma (x)$ and $\Pi (x)$ are defined as: 
\begin{equation}
\label{sp2-curr}
\begin{array}{l}
\Sigma ^\alpha (x)=\dfrac 18\dsum\limits_\mu \left[ 
\overline{\psi }(x)U_\mu (x)t^\alpha \psi (x+\widehat{\mu })+\overline{\psi }%
(x)U_\mu ^{\dagger }(x-\widehat{\mu })t^\alpha \psi (x-\widehat{\mu }%
)\right]  \\ \Pi ^\alpha (x)=\dfrac 18\dsum\limits_\mu \left[ \overline{\psi 
}(x)U_\mu (x)t^\alpha \gamma _5\psi (x+\widehat{\mu })+\overline{\psi }%
(x)U_\mu ^{\dagger }(x-\widehat{\mu })t^\alpha \gamma _5\psi (x-\widehat{\mu 
})\right] 
\end{array}
\end{equation}
and are extended bilinears that, in the continuum limit, reduce to
$S(x)$ and $P(x)$ respectively.

Beside providing the formal invariance of the action with respect to
local chiral transformations, the introduction of the external sources
also offers a powerful tool to compute the correlation functions of the
lattice operators. Indeed, the partition function of the theory:  
\begin{equation}
\label{zetaj}Z(j)=\int \left( dUd\psi d\overline{\psi }\right) \ \exp
\left\{ -S\left( U,\psi ,\overline{\psi };j\right) \right\} 
\end{equation}
is an explicit functional of the external sources $(j=v_\mu, a_\mu, s,
p, \sigma , \pi)$. We can then use eqs.~(\ref{va-def}) and
(\ref{sp-def}) to compute, with standard techniques, the correlation
functions of $V_\mu $, $A_\mu $, $S$ and $P$ as functional derivatives
of $Z(j)$ with respect to the proper external sources%
\footnote{Contrary to what happens in the continuum formalism, however,
here one has to be aware of the fact that the action $S_{\psi j}$ is not
simply linear in the external sources. Thus, the derivatives of the
partition function might differ, from the corresponding correlation
functions, for the contributions of additional contact
terms.}. Eventually, this approach becomes of practical utility once the
partition function is expressed in terms of an effective Lagrangian. In
the following, we discuss how this effective Lagrangian can, in fact, be
calculated. 

\section{The effective lattice Lagrangian in the strong coupling and large $N
$ limit\label{sec:eff-action}}

In the limit of strong coupling and large number of colors $N$, the 
high-energy degrees of freedom of lattice QCD can be analytically
integrated out \cite{klu,ks}. After this integration, the partition
function turns out to be expressed in terms of an effective action,
containing only mesonic fields and external sources. The purpose of this
paper is to investigate the feasibility of performing the same
integration, numerically, in the region of intermediate couplings
relevant for simulations of lattice QCD. Still, the study of the strong
coupling limit is an important preliminary step. The effective lattice
Lagrangian, obtained in this limit, has a general structure, in terms of
effective fields and external sources, that only follows from chiral
invariance, charge conjugation and discrete lattice space-time
symmetries. Consequently, the same structure must be preserved in the
region of intermediate gauge coupling. In this region, one can
consider an effective lattice Lagrangian of this form and determine its
free parameters through a numerical matching with the full QCD lattice
theory.

The procedure to integrate out the high-energy degrees of freedom of
lattice QCD, in the strong coupling and large $N$ limit, has been
developed in refs.~\cite{klu,ks}. In ref.~\cite{cs} the method has been
implemented in the formalism of modern chiral Lagrangian introduced by
Gasser and Leutwyler \cite{gl}. In this section, we apply the results of
these papers to obtain the effective Lagrangian for lattice QCD with
Wilson fermions and the set of external sources introduced in the
previous section.

At the leading order in the strong coupling and large $N$ expansion, the
generating functional of eq.~(\ref{zetaj}) can be expressed in terms of
an effective action \cite{klu,ks}:
\begin{equation}
\label{zeff} 
Z(j)=\int \left( d\cal M \right) \, \exp \left\{ -S_{eff} \left( {\cal
  M} ; j \right) \right\}
\end{equation}
where $S_{eff}({\cal M};j)$ is given by:
\begin{equation}
\label{smes}
S_{eff}({\cal M};j)=-N\ \left\{ \sum_{x,\mu }{\rm Tr}\left[ F\left(
\lambda _\mu (x)\right) \right] -\sum_x{\rm Tr}\left[ \ln {\cal M}(x)\right]
+\sum_x{\rm Tr}\left[ J(x)\,{\cal M}(x)\right] \right\} 
\end{equation}
and ${\cal M}(x)$ is a matrix in the spin and flavour space representing 
an effective bosonic field. The function $F(\lambda )$ in
eq.~(\ref{smes}) is defined as: 
\begin{equation}
\label{effe}F(\lambda )=1-\sqrt{1-\lambda }+\ln \left( \frac{1+\sqrt{%
1-\lambda }}2\right) 
\end{equation}
and the matrix $\lambda _\mu (x)$ is a function of the effective field and the
external sources: 
\begin{equation}
\label{lambdam}\lambda _\mu (x)=-{\cal M}(x)\left[ rK(x)-\gamma _\mu \right]
Y_\mu (x){\cal M}(x+\widehat{\mu })\left[ rK(x+\widehat{\mu })+\gamma _\mu
\right] Y_\mu ^{\dagger }(x) 
\end{equation}
The external sources $J(x)$, $K(x)$ and $Y_\mu(x)$ appearing in
eqs.~(\ref{smes}) and (\ref{lambdam}) have been defined in
eqs.~(\ref{Spsij}) and (\ref{sources}).

The functional integration in eq.~(\ref{zeff}) is over a generalized
contour in the complex matrix space, which can be parameterized by
writing $\cal M$ in polar form: $\cal M = R \, \cal U$, with $\cal R$
and $\cal U$ hermitian and unitary matrices respectively. The functional
measure in eq.~(\ref{zeff}) is $\left( d\cal M \right) = \left( d\, \cal
U \right)$, where $\left( d\, \cal U \right)$ is the Haar measure on the
unitary group of $\cal U$. The functional integral $Z(j)$ is independent
on the choice of $\cal R$. 

In order to evaluate the partition function in perturbation theory,
following refs.~\cite{klu,ks} we look for a translationally invariant
stationary point of the action $S_{eff}({\cal M})$, taken in the the
limit of vanishing external sources. The expansion of $\cal M$ around
its vacuum expectation value will then generate the standard
perturbative expansion and, for large $N$, this also corresponds to
evaluate the partition function $Z(j)$ by the saddle-point method.

The vacuum expectation value of $\cal M$ is found by choosing:
\begin{equation}
\label{saddle}{\cal M}_0(x)=u_0 I
\end{equation}
and considering the effective action in the limit of vanishing external
sources. In this limit, the matrix $\lambda _\mu (x)$ of
eq.~(\ref{lambdam}) reduces to: 
\begin{equation}
\label{lambda0}\lambda _0=(1-r^2)\,u_0^2 
\end{equation}
The stationary point of $S({\cal M})$ is then given by $u_0$ satisfying
the equation: 
\begin{equation}
\label{equ0}\frac{4\,(1-r^2)\,u_0^2}{1+\sqrt{1-(1-r^2)\,u_0^2}}\ +\ \left(
m+4r\right) \,u_0=1 
\end{equation}
whose positive solution is: 
\begin{equation}
\label{u0}u_0(m,r)=\frac{-3\,\left( m+4r\right) \ +\ 4\,\sqrt{\left(
m+4r\right) ^2\,+\,7\,(1-r^2)}}{\left( m+4r\right) ^2\,+\,16\,(1-r^2)} 
\end{equation}
For $r=1$, one simply has: 
\begin{equation}
\label{u0r1}u_0(m,1)=\frac 1{m+4} 
\end{equation}

The matrix ${\cal M}$ can be now expressed in terms of fields with
vanishing vacuum expectation value. A convenient general form is:
\begin{equation}
\label{fullm}{\cal M}(x)=u_0\,\exp \left[ i\,{\cal S}(x)\,+i\,{\cal P}%
(x)\,\gamma _5+i\,{\cal V}_\mu (x)\,\gamma _\mu +i\,{\cal A}_\mu (x)\,\gamma
_\mu \gamma _5+i\,{\cal T}_{\mu \nu }(x)\,\sigma _{\mu \nu }\right] 
\end{equation}
where ${\cal S}$, ${\cal P}$, ${\cal V}_\mu$, ${\cal A}_\mu$ and ${\cal
T}_{\mu \nu}$ are matrices in the pure flavour space. In the lattice
effective theory these fields, which carry integer spin, represent the
mesonic excitations with the corresponding quantum numbers.

The lattice effective action of eq.~(\ref{smes}) differs from the
standard continuum chiral Lagrangian in that the latter is only
expressed in terms of external sources and pseudoscalar fields. Thus, in
order to derive an effective lattice action corresponding to this
continuum Lagrangian, the contributions of the scalar, vector, axial and
tensor components of ${\cal M}$, in eq.~(\ref{fullm}), should be
integrated out. In principle, this could be done explicitly. However, in
the context of the strong coupling approximation considered so far we
are more interested in deriving the general structure of the effective
Lagrangian than in computing its specific expression. Therefore, also in
the attempt to keep discussion simple, we will not perform such an
integration. Rather, we assume a complete decoupling of the higher
resonances (corresponding to the infinite mass limit) by simply
neglecting their contributions. That is, we constrain ${\cal M} (x)$ to
the form:
\begin{equation}
\label{m-u}{\cal M}(x)=u_0\,\exp \left[ i\,{\cal P}(x)\,\gamma _5\right]
=u_0 \left[ U(x)\, \frac{1+\gamma_5}{2}\ +\ U^{\dagger }(x)\, 
\frac{1-\gamma_5}{2} \right] 
\end{equation}
where $U(x)=\exp \left[ i {\cal P}(x)\right]$ is a unitary matrix in the
flavour space. In the continuum limit, $U(x)$ would correspond to the
standard effective field entering in the QCD chiral Lagrangian.

An explicit expression for the effective action, in terms of the field
$U$, is obtained by Taylor expanding the function $F(\lambda )$ of
eq.~(\ref{smes}) around the stationary point $\lambda =\lambda _0$. The
resulting action has then the form of an infinite series: 
\begin{equation}
\label{su} 
S_{eff}(U)= \dsum\limits_{k=1}^\infty S_k(U)\,+\,S_m(U) 
\end{equation}
where:
\begin{equation}
\label{suk} 
S_k(U) = - \frac{NC_k}{4 u_0^{2k}} \ \dsum\limits_{x,\mu } {\rm Tr}\left[ 
\lambda_\mu (x)-\lambda _0 \right]^k 
\end{equation}
and
\begin{equation}
\label{sum} 
S_m(U) = - \frac{NC_m}{2} \ \dsum\limits_x {\rm Tr} \left[ \chi ^{\dagger }
(x)\,U(x)\,+\,\chi (x)\,U^{\dagger }(x)\right]
\end{equation}
The coefficients $C_m$ and $C_k \, (k =1,2,\ldots )$ are given by: 
\begin{equation}
\label{CCC}
C_m=4\,u_0 \qquad ,\qquad C_k=\frac{4\,u_0^{2k}\, F^{(k)}(\lambda _0)}{k!}
\end{equation}
and are of order one in the large $N$ expansion. For practical
calculations, the spin-flavour matrix $\lambda _\mu(x)-\lambda _0$ can
be conveniently written in the form: \renewcommand{\arraystretch}{2.0}
\begin{equation}
\label{lmenol0} 
\begin{array}{rl}
\lambda _\mu (x)-\lambda _0= & -\,\dfrac{u_0^2}{2} \ \left[ 
-\,U(x)\left( \Delta _\mu U(x)\right) ^{\dagger }\left(1+\gamma_5 \right)\,-\,
U^{\dagger}(x)\left(\Delta _\mu U(x)\right) \left(1-\gamma_5 \right)\,+\right. 
\\  & 
+\,r^2U(x)\left( \Omega _\mu ^AU(x)\right) ^{\dagger}\left(1+\gamma_5 \right)\,
+\,r^2U^{\dagger}(x)\left(\Omega _\mu ^BU(x)\right) \left(1 -\gamma_5 \right)\,
+ \\  & \left. 
+\,rU(x)\left( \Sigma _\mu ^AU(x)\right) ^{\dagger }\left(1+\gamma_5 \right)
\gamma_\mu 
+\,rU^{\dagger }(x)\left( \Sigma _\mu ^BU(x)\right) \left( 1-\gamma_5 \right)
\gamma _\mu \right] 
\end{array}
\end{equation}
\renewcommand{\arraystretch}{1.5}where $\Delta _\mu U$ is: 
\begin{equation}
\label{deltamu} 
\Delta _\mu U(x)=W_\mu (x)U(x+\widehat{\mu })Z_\mu ^{\dagger }(x)\ -\ U(x) 
\end{equation}
and the quantities $\Omega _\mu ^{A,B}U$ and $\Sigma _\mu^{A,B}U$ are defined 
as: 
\begin{equation}
\label{operators} 
\begin{array}{l}
\Omega _\mu ^AU(x)=W_\mu (x)\xi (x+\mu )U^{\dagger }(x+\widehat{\mu })W_\mu 
^{\dagger }(x)\xi (x)-\ U(x) \\ 
\Omega _\mu ^BU(x)=\xi (x)Z_\mu (x)U^{\dagger }(x+ \widehat{\mu })\xi (x+\mu )
Z_\mu ^{\dagger }(x)\ -\ U(x) \\ 
\Sigma _\mu ^AU(x)=Z_\mu (x)U^{\dagger }(x+\widehat{\mu })W_\mu ^{\dagger }(x)
\xi (x)-Z_\mu (x)\xi ^{\dagger }(x+\mu)U(x+\widehat{\mu })Z_\mu ^{\dagger }(x) 
\\ \Sigma _\mu ^BU(x)=\xi (x)Z_\mu (x)U^{\dagger }(x+\widehat{\mu })W_\mu 
^{\dagger }(x)-W_\mu (x)U(x+\widehat{\mu })\xi ^{\dagger }(x+\mu )W_\mu 
^{\dagger }(x) 
\end{array}
\end{equation}

Equation (\ref{su}) is the effective chiral Lagrangian for lattice QCD
with Wilson fermions at the leading order in the strong coupling and
large $N$ expansion. In the next section we will study the predictions
of this Lagrangian in some more detail. Here we conclude with some
important remarks.

i) As the original fermionic theory, the effective Lagrangian is
invariant with respect to local chiral transformations. This is easily
shown to be the case when we assume that the field $U(x)$ transforms
according to: 
\begin{equation}
\label{u-t}U(x)\rightarrow R(x)U(x)L^{\dagger }(x) 
\end{equation}
and the external sources $W_\mu$, $Z_\mu$, $\xi$ and $\chi$ have the
transformation properties defined in eq.~(\ref{sources-t}). Notice that
the quantity $\Delta _\mu U(x)$, introduced in eq.~(\ref{deltamu}),
transforms as the matrix $U(x)$ and, in the limit of vanishing external
sources, it reduces to the standard discrete derivative of $U$:
\begin{equation}
\label{dmu}\left( \Delta _\mu U(x)\right) _0=\nabla _\mu U(x)\equiv U(x+ 
\widehat{\mu })-U(x) 
\end{equation}
Thus, $\Delta _\mu U$ represents a possible definition of the lattice
chiral covariant derivative of $U$ and, in the continuum limit, it
reduces in fact to the continuum covariant derivative $D_\mu U$.
 
ii) In the limit of vanishing Wilson term ($r=0$), the quantity
$(\lambda _\mu-\lambda_0)$ in eq.~(\ref{lmenol0}) is proportional to the
covariant derivative $\Delta _\mu U(x)$. Therefore, it vanishes for
vanishing external sources and for constant field $U(x)$. In this limit,
the series expansion of the effective action in eq.~(\ref{su})
corresponds to an expansion in powers of the external momenta $p$, each
term $S_k$ being at least of order $p^k$. As in the continuum case, this
feature is a consequence of chiral symmetry, and is what makes the
continuum QCD chiral Lagrangian a useful tool for the study of
low-energy strong interactions. However, due to the chiral breaking of
the Wilson term, this feature is lost on the lattice. In particular,
contributions of the order of $p^0$ enter in each of the infinite terms
$S_k$ of the effective action. Thus in this case, even in the low-energy
limit, each of these terms should be in principle considered to perform
calculations. Nevertheless, as we will show in the next section, in the
large $N$ limit one can use the standard perturbative technique, so that
the lattice effective action of eq.~(\ref{su}) is still useful in
practice. 

iii) The effective lattice action of eq.~(\ref{su}) is not specifically
a low-energy effective theory. Indeed, but for our deliberate neglecting
of the contributions of the higher resonances, no low-energy
approximation has been introduced to derive it. Rather, eq.~(\ref{su})
defines an effective theory describing, in the strong coupling and large
$N$ limit, the strong interactions among pseudoscalar mesons at
arbitrary energy. In the intermediate coupling region, a corresponding
effective Lagrangian, with the same symmetry properties, can also be
considered. The free parameters entering in this Lagrangian, including
those coming from integrating out the higher resonances, can be
eventually computed by matching with a numerical simulation within the
full lattice theory. In this approach, the infra-red cut-off introduced
by the finite lattice size in numerical simulations does not represent a
problem any longer, since no low-energy expansion has been performed in
deriving the effective theory.

iv) The most evident effect of the Wilson term in the lattice effective
action is the appearance of additional couplings. For $r=0$, the
effective action of eq.~(\ref{su}) reduces to the form:
\begin{equation}
\label{SD}
\begin{array}{l}
S_D(U)=\dfrac 12 N \ \dsum\limits_{x,\mu }{\rm Tr }\left\{ 
\left(C_1+2C_2\right) \, \left[ \left(\Delta _\mu U\right) ^{\dagger }
\left( \Delta _\mu U\right)\right] - C_2 \left[ \left( \Delta _\mu
U\right) ^{\dagger} \left(\Delta _\mu U \right) \right] ^2 \right\}  -
\\ \qquad \qquad - \dfrac 12 N C_m \ \dsum\limits_{x}{\rm Tr }
\left(\chi ^{\dagger}U+ U^{\dagger } \chi \right) + \ldots 
\end{array}
\end{equation}
which, but for the obvious effects of discretization and lattice
breaking of Lorentz invariance, has a direct correspondence with the
continuum chiral Lagrangian.

The presence of the Wilson term in the original lattice action gives
rise to additional interactions in the effective theory, of the kind: 
\begin{equation}
\label{SW}
\begin{array}{l}
S_W(U)=N r^2 \ \dsum\limits_{x,\mu }{\rm Tr }\left\{ 
\dfrac 12 C_1 \left[ U \left(\Omega _\mu ^AU\right) ^{\dagger }+
U^{\dagger }\left( \Omega _\mu ^BU\right) \right] + 
\right. \\ \qquad \qquad
+ \, C_2 \left[ U\left( \Delta _\mu U\right) ^{\dagger}U\left( \Omega 
_\mu ^AU\right) ^{\dagger }\,+\,U^{\dagger }\left( \Delta_\mu U\right) 
U^{\dagger }\left( \Omega _\mu ^BU\right) - 
\right. \\ \qquad \qquad \left. \left.
- \, U\left( \Sigma _\mu ^AU\right)^{\dagger }U^{\dagger }\left( 
\Sigma _\mu ^BU\right) \right] + \ldots \right\}
\end{array}
\end{equation}
These terms are explicit sources of chiral symmetry breaking, since in
the limit of vanishing Wilson sources, $\xi\rightarrow 1$, they reduce
to interactions of the form $U^2$, $\left(\Delta _\mu U\right) \left(
\Delta _\mu U\right)$, \ldots . In the continuum limit ($g\rightarrow
0$), the coefficients of these terms are expected to vanish. However, in
the region of intermediate couplings relevant for current lattice
simulations, their effects are still not negligible. These terms
significantly increase the number of allowed couplings in the lattice
effective theory and, furthermore, they do not have a direct
correspondence in the continuum QCD chiral Lagrangian. In the following
of this paper we will discuss the way to overcome this problem. It will
be shown that an effective lattice theory, without the Wilson terms, can
be in fact considered, and this theory reproduces the values of properly
renormalized lattice correlation functions.

\section{Correlation functions and matrix elements from the effective
Lagrangian \label{sec:CF}}

In this section we use the effective lattice Lagrangian, derived in
eq.~(\ref{su}), to calculate explicitly some simple physical
quantities. From a phenomenological point of view, the results are 
significantly affected by the approximations introduced so far: strong
coupling limit, large $N$ expansion and assumption of complete
decoupling of the higher resonance. On the other hand, these 
calculations achieve a twofold goal: first, they offer us the
opportunity to illustrate the perturbative technique, as applied to the
lattice effective Lagrangian. Possibly, the same technique can be used
for determining the coefficients of the effective Lagrangian in the
intermediate coupling region. In addition, the analysis which follows,
combined with the study of chiral Ward identities in the next section,
will point out some important effects of the several Wilson terms
entering in the lattice effective theory. 

As discussed in sec.~\ref{sec:sources}, the correlation functions of the
lattice operators can be calculated by differentiating the partition
function $Z(j)$ with respect to the external sources. The first
convenient step in order to do that, consists in expressing the
effective field $U(x)$ in terms of $n^2-1$ independent real fields,
representing the true degrees of freedom of the effective theory. We
choose the following parameterization:
\begin{equation}
\label{pgb-field}U(x)=\exp\left\{2\, i\, \phi (x)/F_0\right\} 
\end{equation}
where $\phi (x)=\phi _k(x)\,t_k$ and $F_0$ is a coefficient introduced
to conventionally normalize the pseudoscalar field.

Since the derivatives of the partition functions must be eventually
computed in the limit of vanishing external sources, it is also
convenient to perform, from the beginning, the following change of
variables for the external scalar sources: 
\begin{equation}
\label{new-ss}s_0(x)=\left( m+4r\right) +s_0^{\prime }(x)\qquad ,\qquad
\sigma _0(x)=1+\sigma _0^{\prime }(x) 
\end{equation}
In this way, the new sources $s_0^{\prime }(x)$ and $\sigma _0^{\prime}
(x)$ (we will omit the primes in the following) do actually vanish in
the limit of vanishing external sources.

The next step consists in expanding the effective action (\ref{su}) in  
increasing powers of the pseudoscalar field $\phi$ and the external
sources. Since $\lambda _0$ represents the vacuum expectation value of
the matrix $\lambda _\mu (x)$, the difference $\lambda _\mu
(x)-\lambda_0$ is, by construction, of the order of the field $\phi$ and
the external sources. Thus, although the action (\ref{su}) has the form
of an infinite series in powers of $\left( \lambda _\mu (x)-\lambda
_0\right)$, the above expansion can be systematically performed. For
convenience, we fix the parameter $F_0$ of eq.~(\ref{pgb-field}) in such
a way that the the kinetic term of the effective Lagrangian for the real
scalar field has the standard normalization, i.e. $(1/2)\left( \nabla _\mu
\phi \right) ^2$. In this way we obtain:
\begin{equation}
\label{F0}
F_0^2=2N\left( 1+r^2\right) \left[C_1+2C_2\left( 1-r^2\right) \right]
\end{equation}
where $C_1$ and $C_2$ are the coefficients defined in
eq.~(\ref{CCC}). The effective action $S_\phi$, as a function of the 
field $\phi$ and the external sources, can be written in the form:
\begin{equation}
\label{phi-action}
S_\phi = S_\phi ^{(0)} + S_\phi ^{(1)} + S_\phi ^{(2)} + \ldots   
\end{equation}
where the generic term $S_\phi ^{(k)}$ contains the external sources at
the power $k$. The action $S_\phi ^{(0)}$ represents therefore the
effective lattice action in the limit of vanishing external sources. It
is given by:
\renewcommand{\arraystretch}{2.0}
\begin{equation}
\label{sphi0} 
\begin{array}{l}
S_\phi ^{(0)} = \dsum\limits_x\left\{ 
\dfrac 12 \left(\nabla _\mu \phi \right)^2 \, + \, \dfrac 12 M_\pi ^2 \, 
\phi ^2 \, + \, d_{41} \, K \left(\nabla _\mu \phi \right) ^4 \, + \,
d_{42} \, K \left(\nabla _\mu \phi ^2 \right) ^2 \, + \right. \\ 
\qquad \qquad \left. 
+ \, d_{43} \, K \left(\nabla _\mu \phi ^3 \right) \left(\nabla _\mu \phi 
\right) \, + \, d_{44} \, K \phi ^4 \, + \ldots \, \right\}
\end{array}
\end{equation}
\renewcommand{\arraystretch}{1.5}
For simplicity, all the flavour indices in the above expression have
been omitted. Thus, for instance, $\left(\nabla _\mu \phi \right)^2$
stands for $\left(\nabla _\mu \phi _i \right) \left(\nabla _\mu \phi _
i\right)$ and $K \phi ^4$ for $K_{ijkl} \phi _i \phi _j \phi _k \phi
_l$. The tensor $K_{ijkl}$ is given by:
\begin{equation}
\label{Kijkl} 
K_{ijkl} = \frac 1n \delta _{ij} \delta _{kl} + \frac 12  d_{ijs} d_{kls}  
\end{equation}
and the structure constants $d_{ijk}$, as well as the constants $f_{ijk}$ to 
be introduced below, are defined by the algebra: 
\begin{equation}
\label{algebra}
{\rm Tr} (t_it_j)=\frac 12 \delta _{ij} \quad ,\quad 
\left[ t_i,t_j\right] =i\,f_{ijk}t_k \quad ,\quad 
\left\{t_i,t_j\right\} =\frac 1n\delta _{ij}+d_{ijk}t_k 
\end{equation}
The coefficient $M_\pi^2$ of the $\phi^2$ term in eq.~(\ref{sphi0}) (we
are anticipating that this coefficient represents in fact the
pseudoscalar meson mass) is given by:
\begin{equation}
\label{coeffm} 
M_\pi^2 = \frac{2 N C_m}{F_0^2} \left( m+4r\right) - \frac{16r^2}
{\left( 1+r^2\right)} 
\end{equation}
and the couplings $d_{ij}$ of the several $\phi^4$ terms are:
\renewcommand{\arraystretch}{2.5}
\begin{equation}
\label{coeff0} 
\begin{array}{l}
d_{41}= - \dfrac{\left(1+6r^2+r^4 \right) A_2}{2N\left( 1+r^2 \right)^2 
A_1^2} \quad ,\quad 
d_{42} =  \dfrac{\left(1-r^2 \right) A_1 - 64 r^2 A_2}{8N\left( 1+r^2 \right)
^2 A_1^2} \\ 
d_{43}= - \dfrac{\left(1+r^2 \right) A_1 - 24 r^2 \left( 3 + r^2 \right) A_2}
{6N\left( 1+r^2 \right) ^2 A_1^2} 
\quad ,\quad d_{44} = \dfrac{8r^2 \left(A_1 - 12 r^2 A_2\right)}
{3N\left( 1+r^2 \right)^2 A_1^2} 
\end{array}
\end{equation}
\renewcommand{\arraystretch}{1.5}where:
\begin{equation}
\label{A12} 
\begin{array}{l}
A_1=C_1+2\, C_2\left( 1-r^2\right) \\
A_2=C_2+3\, C_3\left( 1-r^2\right) +2\, C_4\left( 1-r^2\right) ^2
\end{array}
\end{equation}

Since, at the leading order, the effective action $S_{eff}(U)$ is
proportional to the number of colors $N$, the perturbative expansion in
terms of the scalar field $\phi$ corresponds in fact to an expansion in
increasing powers of $1/N$. The couplings of the $\phi^2$ terms are by
construction of order 1 and the couplings of the terms $\phi^{2k}$ are
of order $1/N^{k-1}$ (see eq.~(\ref{coeff0})). Thus, in the large $N$
limit, the lattice effective Lagrangian can be treated in perturbation
theory. Furthermore, it is important to remember that, in this context,
only the lowest (non trivial) perturbative order should be considered in
each particular calculation, since contributions suppressed by
additional powers of $1/N$ have been already neglected, from the very
beginning, in the calculation of the effective Lagrangian itself.

At the leading order in perturbation theory, the propagator of the
scalar field $\phi$ satisfies 
the equation: 
\begin{equation}
\label{giacomo}
\left( \nabla^2 _x -M_\pi^2\right) \Delta (x,y)=-\delta _{x,y} 
\end{equation}
where $\nabla^2$ represents the (symmetric) discretized version of the 
four-dimensional Laplacian operator:
\begin{equation}
\label{giacomino}
\nabla^2 f(x) \equiv \dsum\limits_\mu \left[ f(x+\widehat{\mu })-2f(x) + 
f(x-\widehat{\mu }) \right]
\end{equation}
The solution of eq.~(\ref{giacomo}) is the standard lattice propagator
of the scalar field: 
\begin{equation}
\label{deltaxy}\Delta (x,y)=\frac 1V\dsum\limits_p\frac{e^{-ip(x-y)}}{%
4\dsum\limits_\mu \left[ \sin {}^2(p_\mu /2)\right] +M_\pi^2} 
\end{equation}
where $V=L_x L_y L_z L_t$ is the lattice volume and the sum is extended
over all the lattice momenta, $p_i=2\pi k/L_i$. For later use we also
give the expression of the propagator as a function of the time $t$ and
the spatial momentum $\vec p$. In the infinite volume limit, and for
$t>0$, one finds:
\begin{equation}
\label{deltapt}\Delta (t,\vec p)=\sum_{\vec x}\Delta (t,\vec x;0)\ e^{-i\vec
p\cdot \vec x}=\frac{\ e^{-\widetilde{E}_\pi t}}{2\sinh \widetilde{E}_\pi } 
\end{equation}
where $\widetilde{E}_\pi $ is given by: 
\begin{equation}
\label{enerpi}
\widetilde{E}_\pi =2{\rm arcsinh}\left[ \sin {}^2\left( \frac{p_i}2\right) +
\sinh {}^2\left( \frac{\widetilde{M}_\pi }2\right) \right]^{1/2} 
\end{equation}
and: 
\begin{equation}
\label{masspi}
\widetilde{M}_\pi =2\,{\rm arcsinh}\left( \frac{M_\pi }2\right) 
\end{equation}
In the continuum limit, $\widetilde{M}_\pi \rightarrow M_\pi$ and 
$\widetilde{E}_\pi \rightarrow \sqrt{p_i^2 + M_\pi^2}$.
 
The next term, $S_\phi ^{(1)}$, of the lattice effective action in
eq.~(\ref{phi-action}) is the one linear in the external sources. This
term can be conveniently written in the form:
\begin{equation}
\label{sphi1} 
\begin{array}{l}
S_\phi ^{(1)} =\dsum\limits_x\left\{ 
s_0(x) \widetilde{S}_0(x) + \sigma _0(x) \widetilde{\Sigma}_0 (x) + p_k(x) 
\widetilde{P}_k(x) + \pi _k(x) \widetilde{\Pi}_k (x) + \right. \\
\qquad \qquad \left. + a_\mu ^k(x) \widetilde{A}_\mu ^k(x) + v_\mu ^k(x) 
\widetilde{V}_\mu ^k(x) + \ldots \right\} 
\end{array}
\end{equation}
where the ``effective'' operators $\widetilde{O}_i$ are given by:
\begin{equation}
\label{effop} 
\begin{array}{l}
\widetilde{S}_0 (x) = d_s \, + \, d_{s1} \, \phi _k^2(x) \, + \, \ldots \\
\widetilde{\Sigma}_0 (x) = d_\sigma \, + \, d_{\sigma 1} \, \phi _k^2(x) \, 
+ \, \ldots \\  
\widetilde{P}_k (x) = d_p \, \phi_k(x) \, + \, \ldots \\
\widetilde{\Pi}_k (x) = d_{\pi 1} \, \phi _k(x) \, + \, d_{\pi 2}\, \nabla^2 
\phi_k(x) \, +\, \ldots \\
\widetilde{A}_\mu ^k(x) = d_a \, \nabla _\mu \phi _k(x) \, + \, \ldots \\
\widetilde{V}_\mu ^k(x) = d_v \, f_{ijk} \, \phi _i(x) \left( \nabla _\mu 
\phi_j(x)\right) \, + \, \ldots 
\end{array}
\end{equation}
with coefficients:
\begin{equation}
\label{coeff1} 
\begin{array}{l}
d_s=-n N C_m \quad ,\quad d_{s1}=NC_m/F_0^2 \\
d_p=-NC_m/F_0 \quad ,\quad d_a=-F_0/\left( 1+r^2\right) \quad ,\quad d_v=1 \\
d_\sigma =8nNC_1r^2 \quad ,\quad d_{\sigma 1} = -8 r^2/\left( 1+r^2\right) \\
d_{\pi 1}=8r^2F_0/\left( 1+r^2\right) \quad ,\quad 
d_{\pi 2}=r^2F_0/2\left(1+r^2\right)
\end{array}
\end{equation}
In the effective theory, the operators $\widetilde{O}_i$ can be used to 
calculate directly the correlation functions of the corresponding
operators in the original theory. 

When one needs to compute the correlation functions of higher powers of
the lattice operators (like for instance $A_\mu^2$) then additional
contributions might come from terms containing higher powers of the
external sources (like $a_\mu^2$ and $a_\mu^2 \phi^2$). Examples of such
terms, that will be considered in the study of the lattice Ward
identities, are found in the next term of the effective action
(\ref{phi-action}): 
\begin{equation}
\label{sphi2} 
\begin{array}{l}
S_\phi ^{(2)} = \dsum\limits_x\left\{ 
d_{\pi \pi 1} \, \pi_k^2(x) \, + \, d_{\pi \pi 2} \, \left( \nabla _\mu \pi _k
(x)\right)^2 \, + \, d_{aa} \, a_\mu ^k(x)^2 \, + \right. \\ \qquad \qquad 
\left. + \, d_{\pi v} \, f_{ijk} \left[ \left( \nabla _\mu \phi_i(x)\right) 
\pi _j(x)-\phi _i(x)\left( \nabla _\mu \pi_j(x)\right) \right] v_\mu ^k(x) \, 
+ \right. \\ \qquad \qquad \left. + \, d_{av} \, f_{ijk} \, a_\mu ^i(x)v_\mu 
^j(x) \left[ \phi _k(x+\widehat{\mu })+\phi _k(x) \right] \, + \right. \\ 
\qquad \qquad \left. + \, d_{vv} \, f_{ijk}f_{lmk} \, v_\mu ^i(x) v_\mu ^l(x)
\phi _j(x) \phi _m(x+\widehat{\mu})\, + \,\ldots \right\} 
\end{array}
\end{equation}
where:
\begin{equation}
\label{coeff2} 
\begin{array}{l}
d_{\pi \pi 1}=2NC_1r^2-2r^2F_0^2/\left( 1+r^2\right) \quad ,\quad 
d_{\pi \pi 2}=r^2F_0^2/8\left( 1+r^2\right) \\ 
d_{aa}=-d_{av}=F_0^2/2\left( 1+r^2\right)  \quad ,\quad 
d_{\pi v}=r^2F_0/2\left( 1+r^2\right) \\ d_{vv}=1/2
\end{array}
\end{equation}

In the effective theory, the correlation functions are computed by
expressing the partition function $Z(j)$ in terms of a functional
integral on the fields $\phi$. Up to higher terms in $1/N$, the 
functional measure simply reduces to $\left( d\phi \right)$:   
\begin{equation}
\label{Zchiral}Z(j)=\int \left( d\phi \right) \ \exp \left\{ -S_\phi \left(
\phi ;j\right) \right\} 
\end{equation}
All the correlation functions of interest can be then obtained by
subsequent derivatives of $Z(j)$ with respect to the external
sources. We conclude this section by discussing a few interesting
examples. 

\subsubsection*{The quark condensate}
The simplest relevant case is the calculation of the quark condensate
$\langle \overline{\psi }\psi \rangle $. This quantity vanishes to all
orders in standard perturbation theory and a value of $\langle
\overline{ \psi }\psi \rangle $ different from zero is expected to be
the signal of spontaneous chiral symmetry breaking. In the effective
theory, at the leading order of the strong coupling and large $N$
expansion, the absolute value of the condensate for a single flavour of
quarks is given by: 
\begin{equation}
\label{psipsi}
\langle \overline{\psi }\psi \rangle = \frac 1n \left( \frac
1Z\frac{\delta Z}{\delta s^0(x)}\right) _0=-\frac{d_s} {n}=4Nu_0 
\end{equation}
Thus $u_0$, the vacuum expectation value of the effective field ${\cal
M} (x)$, basically represents the quark condensate.

\subsubsection*{The pion mass and the critical quark mass}
In the effective theory, the correlation function of two pseudoscalar 
densities is given by: 
\begin{equation}
\label{pp}\langle P^j(x)P^{k\dagger }(y)\rangle =\left( \frac 1Z\frac{\delta
^2Z}{\delta p^j(x)\delta p^k(y)}\right) _0=d_p^2\ \delta ^{jk}\,\Delta (x,y) 
\end{equation}
where $\Delta (x,y)$ is the propagator of the $\phi$-field of
eq.~(\ref{deltaxy}). At this order, the pseudoscalar correlation
function only receives contribution from the propagation of a single
pseudoscalar meson. Thus, $\Delta (x,y)$ is just the pion propagator and
the pole of this function,  $M_\pi^2$, defines the pion mass. From
eq.~(\ref{coeffm}) we can write this mass in the form:
\begin{equation}
\label{Mpi-pole} M_\pi ^2=\frac{2NC_m}{F_0^2}\,\left( m-\widehat{m}%
(m,r)\right) 
\end{equation}
where $\widehat{m}(m,r)$ is the function: 
\begin{equation}
\label{m-cap}
\widehat{m}(m,r)=\frac{8\,r^2 F_0^2}{\left( 1+r^2\right) NC_m}-4r 
\end{equation}

The vanishing of the pion mass identifies the chiral limit of the
theory. The corresponding critical value of the quark mass, $m_c$, is
therefore defined by the equation: 
\begin{equation}
\label{mc}m_c=\widehat{m}(m_c,r) 
\end{equation}
In the limit $r=0$, the function $\widehat{m}$ identically vanishes and
the critical quark mass is $m_c=0$. For finite values of $r$, the Wilson
term induces an additive renormalization of the quark mass. In
particular, for $r=1$ one finds from eqs.~(\ref{m-cap}) and (\ref{mc})
that the pion mass vanishes for $m_c=-2$, corresponding to the critical
value $k_c=1/4$ of the Wilson hopping parameter. This is indeed a well
known result of strong coupling lattice QCD%
\footnote{It can be proved, in fact, that the result at $r=1$ is true
for any $N$, and applies equally well to $N=3$ \cite{ks}.}.

In numerical simulations, it is often convenient to compute the
correlation functions as a function of the time $t$ and the spatial
momentum $\vec p$. In the effective theory (in the infinite volume
limit) these functions can be calculated by using
eq.~(\ref{deltapt}). Thus, for instance, for the pseudoscalar
correlation function we find: 
\begin{equation}
\label{cpp}C_{PP}^{jk}(t,\vec p) \equiv \sum_{\vec x}\langle P^j(x)P^{k
\dagger}(0)\rangle e^{-i\vec p\cdot \vec x}=\delta ^{jk\ }\,\frac{\,d_p^2\ 
e^{- \widetilde{E}_\pi t}}{2\sinh \widetilde{E}_\pi } 
\end{equation}
with $\widetilde{E}_\pi$ given by eq.~(\ref{enerpi}). It is interesting
to notice that the energy dependence of the correlation function
predicted by the effective Lagrangian differs, by lattice discretization
effects, from the corresponding continuum form: 
\begin{equation}
\label{c-cont}C_{PP}^{jk}(t,\vec p)\sim \frac{e^{-E_\pi t}}{2E_\pi} 
\end{equation}
where $E_\pi=\left( \vec{p}^{\,2}+M_\pi^2\right)^{1/2}$. A behaviour as
given in eq.~(\ref{cpp}) has been observed in numerical QCD simulations 
\cite{ape-ukqcd}. Therefore, this is an example of pure lattice
artifacts, that are actually found in numerical calculations, and whose
existence is predicted by the lattice effective Lagrangian, even in the
strong coupling approximation considered so far. It is likely that, in
the numerical determination of the coefficients of the effective
Lagrangian in the intermediate coupling region, the possibility of
taking into account such lattice artifacts will help in better
controlling the associated systematic errors. This is another advantage
of introducing the effective Lagrangian on the lattice as an
intermediate step in the calculations. 

\subsubsection*{The pseudoscalar decay constant}

The decay constant of pseudoscalar mesons can be computed from the
following correlation function: 
\begin{equation}
\label{ca0p}
C_{A_0P}^{jk}(t,\vec p)\equiv\sum_{\vec x}\langle A_0^j(x)P^{k\dagger}
(0)\rangle e^{-i\vec p\cdot \vec x}=-\delta ^{jk\ }\,\frac{d_p\,d_a}{2\sinh 
\widetilde{E}_\pi }\,\left( 1-e^{-\widetilde{E}_\pi }\right) \,e^{- 
\widetilde{E}_\pi t} 
\end{equation}
In general, two-point correlation functions in momentum space can be
used to extract the values of the matrix elements of lattice operators
between the vacuum and a single particle state. For instance, the
correlation function of any two operators $O_A$ and $O_B$, carrying the
quantum number of the pion field, are expected to have in the large time
limit the following behaviour%
\footnote{For convenience we choose here, for the one-particle states, 
the normalization: 
\begin{equation} 
\langle M(\vec p)|M(\vec{p}^{\,\prime})\rangle = 2 \sinh \widetilde{E}_M
\delta _{\vec p \vec{p}^\prime} 
\end{equation}
instead of the standard covariant normalization:
\begin{equation} 
\langle M(\vec p)|M(\vec{p}^{\,\prime})\rangle = 2 E_M \delta _{\vec p
\vec{p}^\prime}  
\end{equation}}: 
\begin{equation}
\label{cab}\lim _{t\rightarrow \infty }C_{AB}(t,\vec p)=\frac{\langle
0|O_A(0)|\pi \rangle \langle \pi |O_B^{\dagger }(0)|0\rangle }{2\sinh 
\widetilde{E}_\pi } \ e^{-\widetilde{E}_\pi t} 
\end{equation}
At the lowest order in the strong coupling effective theory the large
time limit is not a necessary requirement, since the correlation
functions are always of the form (\ref{cab}). Thus, from
eqs.~(\ref{cpp}) and (\ref{ca0p}), we can obtain (up to irrelevant 
complex phases) the values of the following matrix elements: 
\begin{equation}
\label{me}\langle 0|P^j(0)|\pi ^k\rangle =\delta ^{jk}d_p \quad , \quad 
\langle 0|A_0^j(0)|\pi ^k \rangle =-\delta ^{jk}d_a\left(1 - e^
{-\widetilde{E}_\pi }\right)
\end{equation}
The last equality provides the value of the pseudoscalar decay constant
$F_\pi$ in the strong coupling limit. By considering the continuum
definition $\langle 0|A_0^j(0)|\pi ^k \rangle =\delta ^{jk}F_\pi E_\pi$,
we find that, up to lattice artifacts:  
\begin{equation}
\label{fpai}F_\pi =-d_a = F_0/(1+r^2) 
\end{equation}
From eqs.~(\ref{psipsi}), (\ref{Mpi-pole}) and (\ref{fpai}), we can also 
derive a relation between $F_\pi$ and the quark condensate: 
\begin{equation}
\label{gell-mann}\left( 1+r^2\right)^2 F_\pi ^2\,M_\pi ^2=2\left( m-
\widehat{m} \right) \langle \overline{\psi }\psi \rangle 
\end{equation}
In the next section, the same relation will be derived directly from the 
axial Ward identity of the lattice theory. In that context the meaning
of the factors $\left( 1+r^2\right)$, entering in eqs.~(\ref{fpai}) and 
(\ref{gell-mann}) because of the Wilson term, will be explicitly
clarified. 

\subsubsection*{The form factor of semileptonic pion decay}

In the limit of exact isotopic invariance, the semileptonic decays of
charged pions, $\pi^+ \rightarrow \pi^0 e^+ \nu_e$, are described in
terms of a single invariant form factor, $f_+(q^2)$. This form factor
parameterizes the matrix element of the weak vector current between
external pion states:
\begin{equation}
\label{formf}
\langle \pi^0 (\vec p _2) \vert \bar u \gamma _\mu d \vert \pi^+ (\vec p _1)
\rangle = \sqrt 2 \, f_+(q^2) \, \left( p_1 + p_2 \right)_\mu 
\end{equation}
$q_\mu=(p_1-p_2)_\mu$ is the momentum transferred in the decay and the
form factor is normalized at $q^2=0$ by current conservation:
$f_+(0)=1$. 

The matrix element of eq.~(\ref{formf}) can be computed by considering
the following three-point correlation function: 
\renewcommand{\arraystretch}{2.0} 
\begin{equation}
\label{cpvp}
\begin{array}{l}
C_{\mu}^{jkl}(t_x,t_y;\vec p_1, \vec p_2) \equiv \dsum\limits_{\vec x, \vec 
y} \langle P^j(y) V_\mu^k (0) P^{l\dagger}(x)\rangle \, e^{i\vec p_1\cdot 
\vec x -i\vec p_2\cdot \vec y} = \\ 
\qquad \qquad = \langle 0|P^j(0)|\pi^r (\vec p_2) \rangle \,
\langle \pi^r (\vec p_2)|V_\mu^k(0)|\pi^s (\vec p_1) \rangle \,
\langle \pi^s (\vec p_1)|P^{l\dagger}(0)|0\rangle \, \cdot \\ 
\qquad \qquad \cdot \dfrac{e^{-\widetilde{E}_2 t_y + \widetilde{E}_1 t_x}}
{4\sinh \widetilde{E}_1 \sinh \widetilde{E}_2} 
\end{array}
\end{equation}
\renewcommand{\arraystretch}{1.5}where the last equality holds for $t_x
< 0 < t_y$. An explicit calculation of this function in the effective
theory shows that the matrix element of the weak current between two
external pion states has the form:
\begin{equation}
\label{mele}
\langle \pi^j (\vec p_2)|V_\mu^k(0)|\pi^l (\vec p_1) \rangle = -i \, d_v \,
f_{jkl} \left(e^{ip_{1\mu}}-e^{-ip_{2\mu}} \right)
\end{equation}
Thus, up to lattice artifacts, we find that the vector form factor for
pion semileptonic decays is just given by:
\begin{equation}
\label{fform}
f_+(q^2) = d_v = 1
\end{equation}

This result has likely a simple explanation. In
reality, the form factor $f_+(q^2)$ turns out to be well described by a
phenomenological vector meson dominance model, where it is expected to
have the form: $f_+ (q^2)=m_\rho/\left( m_\rho-q^2\right)$, with
$m_\rho$ the rho meson mass. In this case, in the limit of complete
decoupling of the rho meson as the one considered so far, the form
factor becomes completely independent on $q^2$ and reduces to 1 for
current conservation. In a typical lattice simulation, the value of
$q^2/m_\rho^2$ is roughly of the order of 1. Thus, in that context, the
assumption of complete decoupling of the rho meson is certainly
unreliable. What we expect to happen is that the effect of the rho and
other resonances will manifest itself through the structure of the
couplings in the effective lattice Lagrangian, which, rather than being
limited to the nearest neighbors as in the strong coupling
approximation, will fall off with a range $\approx m_\rho^{-1}$. It
should be also noticed that the equality $d_v=1$, obtained for the
coefficient of the effective theory, is a consequence of the lattice
vector current conservation. In the next section we will find in fact
that the same equality can be also obtained directly from the vector
Ward identities of the lattice theory.
 
\subsubsection*{The $B_K$-parameter}

The calculation of the $B_K$-parameter in the effective theory shows
that some chiral symmetry breaking effects, induced on the lattice by
the Wilson term \cite{bochicchio,bkt} and actually observed in numerical
simulations (see e.g. ref.~\cite{bkn}), do not appear in the strong
coupling and large 
$N$ limit. This seems to be a quite general property of the theory in
this limit: that is, the chiral symmetry is not broken to all the
extents it could possibly be. The same feature will be illustrated by
the study of the lattice Ward  identities in the next section. 

The parameter $B_K$ parameterizes the matrix element $\langle
\overline{K} ^0 \vert O^{\Delta S=2} \vert K^0 \rangle$ relevant for
$K^0 -\overline{K}^0$ mixing in weak interactions. $O^{\Delta S=2}$ is
the four-fermion operator $\left (\overline s \gamma _\mu ( 1-\gamma_5)
d \right)^2$ entering in the effective $\Delta S=2$ weak
Hamiltonian. For external states of kaons at rest, the parameter $B_K$ 
is defined as: 
\begin{equation}
\label{bk1} 
\langle \overline{K} ^0 \vert O^{\Delta S=2} \vert K^0 \rangle = \frac{16}{3}
B_K F_K^2 M_K^2
\end{equation}
This definition is such that in the vacuum saturation approximation:
\begin{equation}
\label{via} 
B_K=\frac 34 \left( 1+ \frac 1N \right)
\end{equation}
so that, in this limit, $B_K=1$ for $N=3$. 

On the lattice, with Wilson fermions, the matrix element between
external states of arbitrary momenta can be conveniently parameterized
in the form: 
\begin{equation}
\label{bkme} 
\langle \overline{K} ^0 (\vec{p}_2) \vert O^{\Delta S=2} \vert K^0
(\vec{p}_1) \rangle = \alpha + \beta M_K^2 + \gamma \left(p_1 \cdot
p_2\right) + \ldots 
\end{equation}
where the dots stand for higher order terms in $M_K^2$ and $\left(p_1
\cdot p_2\right)$. The parameters $\alpha$ and $\beta$ are lattice
artifacts that should vanish in the true continuum limit, $g\rightarrow
0$. The origin of these terms is the mixing of the lattice operator
$O^{\Delta S=2}$ with operators of different chirality \cite{bkt}, and
it is only due to the presence of the Wilson term in the lattice action.

The calculation of the matrix element (\ref{bkme}) in the effective
theory proceeds through the evaluation of the corresponding three-point
correlation function, by using the technique discussed above. In the
large $N$ limit, one finds that the matrix element only receives
contribution from the $A^2$ part of the $(V-A)^2$ lattice operator. The
matrix elements of the $VA$ and $AV$ operators vanish identically for
parity invariance and the remaining contribution, $V^2$, is suppressed
by an additional power of $1/N$. Technically, this follows from the fact
that the relevant coefficients in the effective Lagrangian are $d_a \sim
\sqrt N$ and $d_v=1$ (see eq.~(\ref{coeff1})). In addition, we find that
in the large $N$ limit the matrix element is saturated by the vacuum
insertion and it is given by:
\begin{equation}
\label{bklme} 
\langle \overline{K} ^0 (\vec{p}_2) \vert O^{\Delta S=2} \vert K^0
(\vec{p}_1) \rangle = 4 \, d_a^2 \left(e^{ip_{1\mu}}-1 \right)
\left(e^{-ip_{2\mu}} -1 
\right) 
\end{equation}
Up to discretization effects, eq.~(\ref{bklme}) corresponds to
eq.~(\ref{bkme}) with $\alpha=\beta=0$, as would be predicted by
unbroken chiral symmetry. Therefore, at the leading order of strong
coupling and large $N$ expansion, the mixing of $O^{\Delta S=2}$ with
operators of different chirality does not occur. We also find in
this limit $\gamma = 4 \, d_a^2 = 4 F_K^2$, corresponding to:
\begin{equation}
\label{bk2} 
B_K = \frac 3 4
\end{equation}
According to eq.~(\ref{via}), this is the result predicted by the vacuum
insertion approximation in the large $N$ limit.

Although the $1/N$ corrections to eq.~(\ref{bk2}) cannot be consistently
computed within the approximations considered so far, it may be useful
to observe that $1/N$ suppressed contributions to the parameter $\alpha$
in eq.~(\ref{bkme}) in fact exist. They come for instance from the
Wilson term:
\begin{equation}
\label{alfa}
-N C_1 r^2 \sum\limits_{x,\mu} {\rm Tr} \left[ a_\mu(x) \phi(x+\widehat{\mu})
a_\mu(x) \phi(x) + \left( a_\mu \rightarrow v_\mu \right)\right]
\end{equation}
which can be obtained by expanding the effective action $S_1$ of
eq.~(\ref{su})%
\footnote{The same combination (\ref{alfa}) is also found in the
symmetric term $\left(\Delta_\mu U \right)^{\dagger} \left(\Delta_\mu U
\right)$ of the effective action, but now with a relative minus sign
between the vector and axial part. Thus, in this case, the contributions
of the two terms cancel in the final result.}. 
Other $1/N$ suppressed contributions to $\alpha$ are also found at the
next order in perturbation theory, coming from the pure $\phi^4$ term of
the effective action (see eq.~(\ref{sphi0})). Since the coefficient
$d_{44}$ of this term is proportional to $r^2$, it represents again a
pure effect of the Wilson chiral symmetry breaking. 

\section{Chiral invariance and Ward identities\label{sec:ward}}

The chiral invariance of the Wilson action, in the presence of external 
sources, implies the existence of a full set of vector and axial Ward
identities (WI). These identities have been studied in
ref.~\cite{bochicchio} to show that, in the continuum limit, they reduce
to the known WI's of QCD. Precisely, once the chiral limit of the theory
is correctly identified and the lattice operators are properly
renormalized, the WI's on the lattice reproduce the relations of
continuum current algebra.

In this section, we will first verify that the lattice WI's are indeed
satisfied, by explicitly computing the relevant correlation functions in
the effective theory. This is nothing more than a useful check of the
calculations performed so far. Then, however, by using these results, we
will show that the relations of continuum current algebra are exactly
reproduced by the lattice theory even in the (leading order) strong
coupling and large $N$ expansion considered so far. In a sense, this is
quite a unexpected result. According to ref.~\cite{bochicchio}, the
current algebra relations are expected to be recovered only in the
continuum limit, when the contribution of higher dimensional operators
to the correlation functions of the WI's can be exactly neglected. In
contrast, the strong coupling approximation represents rather the
opposite limit, in which such contributions, if any, should appear as
genuine and finite corrections. 

For the purpose of this paper, the possibility of recovering exact
chiral symmetry with the Wilson action in the strong coupling
approximation is a very useful result. In fact we can define, even in
the strong coupling limit, ``renormalized'' operators that satisfy all
the WI's of the continuum theory. Then in the next section, by
considering some specific examples, we will be able to show,
analytically, that the correlation functions of these renormalized
operators are completely reproduced by a lattice effective Lagrangian
which does not contain the Wilson terms at all. Thus, each term of this
Lagrangian, but for discretization effects, has a well defined
correspondence in the continuum effective theory. The same result must
be valid in the scaling region of couplings relevant for QCD
simulations. In that case, this effective lattice Lagrangian can be
eventually expanded in powers of the lattice spacing and external
momenta to compute the coefficients of the continuum QCD chiral
Lagrangian.

In order to study the lattice WI's in the framework of the effective
theory, it is convenient to derive them in the form of equalities among
derivatives of the partition function with respect to the external
sources (see footnote 1). 

Let us consider the set of infinitesimal vector and axial
transformations generated, from eqs.~(\ref{psi-t}) and
(\ref{sources-t}), in the limit in which: 
\begin{equation}
\label{inf-trasf} 
\begin{array}{c}
R(x)\simeq 1+i\left[ \alpha _V(x)+\alpha _A(x)\right] \\ 
L(x)\simeq 1+i\left[ \alpha _V(x)-\alpha _A(x)\right] 
\end{array}
\end{equation}
where $\alpha _{V,A}(x)= \alpha _{V,A}^k(x)\,t^k$ are infinitesimal 
quantities. In this limit, the chiral transformations of the quark
fields in eq.~(\ref{psi-t}) reduce to: 
\begin{equation}
\label{psi-var} 
\begin{array}{l}
\psi (x)\rightarrow \,\psi ^{\prime }(x)=\psi (x)+i\,\left[ \alpha
_V(x)+\alpha _A(x)\gamma _5\right] \,\psi (x) \\ 
\overline{\psi }(x)\rightarrow \,\overline{\psi }^{\prime }(x)=\,\overline{%
\psi }(x)-i\,\overline{\psi }(x)\,\left[ \alpha _V(x)-\gamma _5\alpha
_A(x)\right] 
\end{array}
\end{equation}
Similarly, from eq.~(\ref{sources-t}), we can derive the transformations
of the external sources. For the vector and axial sources they have the
form:
\begin{equation}
\label{sources-var1} 
\begin{array}{l}
\delta v_\mu (x)=\nabla _\mu \alpha _V(x)+\dfrac i2\left[ \alpha _V(x+ 
\widehat{\mu })+\alpha _V(x),v_\mu (x)\right] + \\ \qquad \qquad +\dfrac
i2\left[ \alpha _A(x+ 
\widehat{\mu })+\alpha _A(x),a_\mu (x)\right] + \ldots \\ \delta a_\mu
(x)=\nabla _\mu \alpha _A(x)+\dfrac i2\left[ \alpha _V(x+ 
\widehat{\mu })+\alpha _V(x),a_\mu (x)\right] + \\ \qquad \qquad +\dfrac
i2\left[ \alpha _A(x+ 
\widehat{\mu })+\alpha _A(x),v_\mu (x)\right] +\ldots
\end{array}
\end{equation}
where the dots indicate terms of higher order in the external
sources. For the scalar and pseudoscalar sources these transformations,
by being local, have the same form expected in the continuum theory:
\begin{equation}
\label{sources-var2} 
\begin{array}{l}
\delta \sigma (x)=\ i\left[ \alpha _V(x),\sigma (x)\right] -\ \left\{ \alpha 
_A(x),\pi (x)\right\} \\ 
\delta \pi (x)=\ i\left[ \alpha _V(x),\pi (x)\right] +\ \left\{ \alpha
_A(x),\sigma (x)\right\} \\ 
\delta s(x)=\ i\left[ \alpha _V(x),s(x)\right] -\ \left\{ \alpha
_A(x),p(x)\right\} \\ 
\delta p(x)=\ i\left[ \alpha _V(x),p(x)\right] +\ \left\{ \alpha
_A(x),s(x)\right\} 
\end{array}
\end{equation}
Let us then perform in the partition function $Z(j)$ of the full lattice
theory the change of variables defined by eq.~(\ref{psi-var}). With
respect to this change of variables the measure in the functional
integral is invariant%
\footnote{In the more general case of $U(n)\otimes U(n)$ flavour
transformations the invariance of the measure is lost and the consequent
variation generates the quantum anomaly.}. 
On the other hand, by definition, a change of variables does not affect
the value of the integral. Thus, by considering that the lattice action
is invariant with respect to combined transformations of the quark
fields and external sources, it is easy to proof that the partition
function itself is invariant with respect to transformations of the
external sources only. Indeed, by denoting with a prime the transformed
quantities, one has:
\begin{equation}
\label{inv} 
\begin{array}{ll}
Z(j) & =
\dint \left( dU d\psi d\overline{\psi }\right) \ \exp \left\{ -S\left(
U, \psi ,\overline{\psi } ; j\right) \right\} = \\ 
& =\dint \left( dU d\psi d\overline{\psi }\right) \ \exp \left\{ -S\left(
U, \psi^{\prime} ,\overline{\psi^{\prime}} ; j^{\prime} \right) \right\} = \\  
& =\dint \left( dU d\psi ^{\prime } d\overline{\psi^{\prime }}\right) \
\exp \left\{ -S\left( U,\psi ^{\prime }, \overline{\psi^{\prime }};
j^{\prime }\right) \right\} =Z(j^{\prime})  
\end{array}
\end{equation}
Formally, we can express this invariance in infinitesimal form:
\begin{equation}
\label{varz}\delta Z(j)=\sum_a\sum_x\left( \frac{\delta Z(j)}{\delta j_a(x)}%
\right) \delta j_a(x)=0 
\end{equation}
where $\delta j(x)$ are the variations of the sources given in
eqs.~(\ref{sources-var1}) and (\ref{sources-var2}). The whole set of
vector and axial WI's are finally obtained by further differentiating
the above equation with respect to the external sources and taking the
limit of vanishing external sources. For example, for the two- and
three-point correlations functions, the generic WI's have the form: 
\begin{equation}
\label{ward2}\sum_a\sum_x\left[ \left( \frac{\delta ^2Z(j)}{\delta j_a(x)\
\delta j_b(y)}\right) \delta j_a(x)+\left( \frac{\delta Z(j)}{\delta j_a(x)}%
\right) \left( \frac{\delta j_a(x)}{\delta j_b(y)}\right) \right] _0=0 
\end{equation}
and \renewcommand{\arraystretch}{2.5} 
\begin{equation}
\label{ward3} 
\begin{array}{l}
\dsum\limits_a\dsum\limits_x\left[ \left( 
\dfrac{\delta ^3Z(j)}{\delta j_a(x)\ \delta j_b(y)\ \delta j_c(z)}\right)
\delta j_a(x)+\left( \dfrac{\delta ^2Z(j)}{\delta j_a(x)\ \delta j_b(y)}%
\right) \left( \dfrac{\delta j_a(x)}{\delta j_c(z)}\right) +\right. \\ 
\qquad \;\left. +\left( \dfrac{\delta ^2Z(j)}{\delta j_a(x)\ \delta j_c(z)}%
\right) \left( \dfrac{\delta j_a(x)}{\delta j_b(y)}\right) +\left( \dfrac{%
\delta Z(j)}{\delta j_a(x)}\right) \left( \dfrac{\delta ^2j_a(x)}{\delta
j_c(z)\ \delta j_b(y)}\right) \right] _0=0 
\end{array}
\end{equation}
\renewcommand{\arraystretch}{1.5}respectively.

To begin the analysis of the WI's in the effective theory, let us
consider the simple case of the two-point axial-pseudoscalar correlation
function. The corresponding WI is obtained from eq.~(\ref{ward2}) by
choosing $j^b(y)=p^j (y)$. The result reads:
\renewcommand{\arraystretch}{2.5} 
\begin{equation}
\label{ap-wi1}
\begin{array}{l}
\nabla _\mu ^{Lx}\left( \dfrac 1Z \dfrac{\delta ^2Z}{\delta a_\mu ^k(x)
\delta p^j(y)}\right) _0=2\,\left(m+4r\right) \, \left( \dfrac 1Z\dfrac{
\delta ^2Z}{\delta p^k(x)\delta p^j(y)} \right) _0+ \\ \qquad \qquad +\,
2\, \left( \dfrac 1Z\dfrac{\delta ^2Z}{\delta \pi ^k(x)\delta p^j(y)}
\right) _0-\dfrac 1n\ \delta ^{kj}\,\delta _{x,y}\, \left( \dfrac 1Z
\dfrac{ \delta Z}{\delta s^0(x)}\right) _0
\end{array}
\end{equation}
\renewcommand{\arraystretch}{1.5}where $\nabla _\mu^L$ denotes the left 
discrete derivative: 
\begin{equation}
\label{dmu-left}\nabla _\mu ^Lf(x) \equiv f(x)-f(x-\widehat{\mu }) 
\end{equation}
and, we recall, $n$ in eq.~(\ref{ap-wi1}) is the number of flavours.

In the effective theory, the derivatives of the partition function with
respect to the external sources can be easily calculated . For the terms 
entering in eq.~(\ref{ap-wi1}) we find, at the lowest order:
\renewcommand{\arraystretch}{2.5} 
\begin{equation}
\label{ap-wi-cf} 
\begin{array}{l}
\nabla _\mu ^{Lx}\left( \dfrac 1Z 
\dfrac{\delta ^2Z}{\delta a_\mu ^k(x)\delta p^j(y)}\right) _0=d_ad_pM_\pi
^2\,\delta ^{kj}\,\Delta (x,y)-d_ad_p\,\delta ^{kj}\delta _{x,y} \\ \left(
\dfrac 1Z 
\dfrac{\delta ^2Z}{\delta \pi ^k(x)\delta p^j(y)}\right) _0=d_p\left( d_{\pi
1}+d_{\pi 2}M_\pi ^2\right) \,\delta ^{kj}\,\Delta (x,y)-d_pd_{\pi
2}\,\delta ^{kj}\delta _{x,y} \\ \left( \dfrac 1Z\dfrac{\delta ^2Z}{\delta
p^k(x)\delta p^j(y)}\right) _0=d_p^2\,\delta ^{kj}\,\Delta (x,y)\qquad
,\qquad \left( \dfrac 1Z\dfrac{\delta Z}{\delta s^0(x)}\right) _0=-d_s 
\end{array}
\end{equation}
\renewcommand{\arraystretch}{1.5}Now, for the WI (\ref{ap-wi1}) to be 
satisfied, the terms proportional to $\Delta (x,y)$ and those
proportional to the $\delta _{x,y}$ must be equal on both sides of the
equation. From these requirements we find two relations among the
several coefficients $d_i$: 
\begin{equation}
\label{master}\left( d_a-2d_{\pi 2}\right) M_\pi ^2=2\left( m+4r\right)
d_p+2d_{\pi 1} 
\end{equation}
and 
\begin{equation}
\label{ct-ap}\left( d_a-2d_{\pi 2}\right) d_p=-d_s/n 
\end{equation}
By using the explicit values of the coefficients, (eqs.~(\ref{coeff0})
and (\ref{coeff1})), it is easy to verify that both the relations
(\ref{master}) and (\ref{ct-ap}) are indeed satisfied. Thus, in the
effective theory, the axial WI (\ref{ap-wi1}) is identically satisfied
as well. 

To extend this check, we have considered a larger set of both vector and
axial WI's, for two- and three-point correlation functions. In each
case, one or more relations among the coefficients $d_i$ are found, each
of the kind of eq.~(\ref{ct-ap}). Examples of such relations, as derived
from the axial WI's for the correlation functions $\langle A\Pi\rangle$,
$\langle AA\rangle$, $\langle ASP\rangle$ and $\langle AAV\rangle$, are:
\begin{equation}
\label{contacta} 
\begin{array}{lcl}
\left( d_a-2d_{\pi 2}\right) d_{\pi 1}=-d_\sigma /n+4d_{\pi \pi 1} 
\qquad & , & \qquad
\left( d_a-2d_{\pi 2}\right) d_{\pi 2}=-4d_{\pi \pi 2} \\ 
\left( d_a-2d_{\pi 2}\right) d_a=2d_{aa} 
\qquad & , & \qquad
\left( d_a-2d_{\pi 2}\right) d_{s1}=d_p \\ 
\left( d_a-2d_{\pi 2}\right) d_v=2d_{av}-2d_{\pi v}
\qquad & , & \qquad
\left( d_a-2d_{\pi 2}\right) d_v=d_a-2d_{\pi v} 
\end{array}
\end{equation}
Similarly, by considering the vector WI's for $\langle VV\rangle$,
$\langle VPP\rangle$, $\langle VP\Pi\rangle$ and $\langle VAA\rangle$,
we find: 
\begin{equation}
\label{contactv} 
d_v=1 \quad , \quad d_v^2 = 2 d_{vv} \quad , \quad
d_{\pi v} = d_{\pi 2} \quad , \quad d_{a v} = d_a/2
\end{equation}
All the relations in eqs.~(\ref{contacta}) and (\ref{contactv}) are
satisfied regardless the specific values of the coefficients $C_m$,
$C_1$, $C_2$, $\ldots$ entering in the effective Lagrangian. In other
words, eqs.~(\ref{contacta}) and (\ref{contactv}) are a pure consequence
of chiral symmetry and the fact that these relations are satisfied is
basically just a useful check of the calculations performed so far. 

It is useful to rewrite eq.~(\ref{master}) in the following form: 
\begin{equation}
\label{mas2}\left( d_a-2d_{\pi 2}\right) M_\pi ^2=2\left[ m-\widehat{m}%
\left( m,r\right) \right] d_p 
\end{equation}
where we have defined: 
\begin{equation}
\label{m-cap2}\widehat{m}\left( m,r\right) =-\frac{d_{\pi 1}}{d_p}-4r 
\end{equation}
We easily recognize that the identity (\ref{mas2}) is just
eq.~(\ref{Mpi-pole}), that fixes the ratio between the square of the
pion mass and the quark mass. In particular, the function $\widehat{m}$,
defined in eq.~(\ref{m-cap2}), is exactly the same quantity previously
introduced in eq.~(\ref{m-cap}). In addition, the identities
(\ref{mas2}) and (\ref{ct-ap}) can be combined together to obtain: 
\begin{equation}
\label{bobo}\left( d_a-2d_{\pi 2}\right) ^2M_\pi ^2=-2\left[ m-\widehat{m}%
\left( m,r\right) \right] d_s/n 
\end{equation}
By observing that the ratio $\left( d_a-2d_{\pi 2}\right) /d_a$ is equal
to $\left( 1+r^2\right) $, we can also express eq.~(\ref{bobo}) in terms
of the pion decay constant $F_\pi =-d_a$ and the quark condensate
$\langle \overline{\psi} \psi \rangle =-d_s/n$. In this way we obtain: 
\begin{equation}
\label{gell-mann2}\left( 1+r^2\right)^2 F_\pi ^2\,M_\pi ^2=2\left( m-
\widehat{m}\right) \langle \overline{\psi }\psi \rangle 
\end{equation}
that is again eq.~(\ref{gell-mann}) derived before. Thus this equation,
as its continuum analogous, is a direct consequence of the axial WI.

We now show how the continuum current algebra relations and the partial
conservation of the axial current are reproduced in the lattice theory
at the leading order of the strong coupling and large $N$
expansion. Technically, the mechanism turns out to be exactly the same
discussed in ref.~\cite{bochicchio} for the weak coupling regime.

In the continuum theory, the axial WI, expressed in terms of
renormalized operators $O_R(x)$ and the renormalized quark mass $m_R$
has the form: 
\begin{equation}
\label{ap-wic}\partial _\mu ^x\langle A_{R\mu}^k(x) P_R^j(y)\rangle =2\,m_R
\, \langle P_R^k(x) P_R^j(y)\rangle -\dfrac 1n\ \delta ^{kj}\,\delta (x-y)\,
\langle S_R^0(x)\rangle 
\end{equation}
On the lattice, since the fermionic action is simply linear in the
external source $p(x)$, the derivatives of the partition function
entering in the WI (\ref{ap-wi1}) are proportional to the corresponding
correlation functions. Thus, eq.~(\ref{ap-wi1}) can be also written in
the form: 
\begin{equation}
\label{ap-wi2}
\begin{array}{l}
\nabla _\mu ^{Lx}\langle A_\mu ^k(x)P^j(y)\rangle =2\,\left( m+4r\right) \
\langle P^k(x)P^j(y)\rangle -8r\ \langle \Pi ^k(x)P^j(y)\rangle - \\ 
\qquad \qquad \qquad \qquad -\dfrac 1n\ \delta ^{kj}\,\delta _{x,y}\ \langle
S^0(x)\rangle 
\end{array}
\end{equation}
Formally, this differs from the continuum expression (\ref{ap-wic}) for
the presence of terms proportional to $r$, originated by the Wilson term
in the lattice action. We now observe that these contributions can be
expressed in terms of correlation functions of the two lattice operators
$\nabla _\mu^L A_\mu(x)$ and $P(x)$ respectively%
\footnote{In the weak coupling region this is also true, but for the
presence of additional Schwinger terms. This follows from the fact that
$\nabla _\mu^L A_\mu(x)$ and $P(x)$ are the only operators of dimension
four or less with which the Wilson terms can mix \cite{bochicchio}.}. 
Explicitly, from eq.~(\ref{ap-wi-cf}), we find: 
\begin{equation}
\label{boch}
\begin{array}{l}
8r \langle \left[ P^k(x)-\Pi ^k(x)\right] P^j(y)\rangle = \\
\qquad \qquad = \left(1-Z_A \right) \ \nabla _\mu ^{Lx}\langle A_\mu ^k(x)
P^j(y) \rangle -2\, \widehat{m}\,\langle P^k(x)P^j(y)\rangle 
\end{array}
\end{equation}
where:
\begin{equation}
\label{za}
Z_A = 1-2\,\frac{d_{\pi 2}}{d_a}=1+r^2
\end{equation}
Therefore, by substituting eq.~(\ref{boch}) in eq.~(\ref{ap-wi2}), we obtain 
for the lattice WI the following expression: 
\begin{equation}
\label{ap-wi4} Z_A \nabla _\mu ^{Lx}\langle A_\mu^k(x)P^j(y)\rangle =
2\,\left( m-\widehat{m}\right) \langle P^k(x)P^j(y)\rangle -\dfrac 1n \delta 
^{kj}\delta _{x,y} \langle S^0(x)\rangle 
\end{equation}
Despite the strong coupling limit we are considering, this result has
exactly the form (\ref{ap-wic}) required by continuum current algebra,
provided we identify the following renormalized operators and quark
mass: 
\begin{equation}
\label{renorm}
\begin{array}{lll}
A_{R\mu}^k(x)=Z_A A_{\mu}^k(x) \qquad & , & \qquad P_R^k(x)=Z_P P^k(x) \\ 
S_R^0(x)=Z_P S^0(x) \qquad & , & \qquad m_{R}= Z_m\left( m-\widehat{m}
\right) 
\end{array}
\end{equation}
with $Z_A$ given by eq.~(\ref{za}) and $Z_m = 1/Z_P$. In particular, the
vanishing of the renormalized quark mass $m_R$ identifies the proper
chiral limit, in which the axial current becomes exactly conserved and
the pseudoscalar meson mass vanishes.

To verify the extent to which chiral symmetry is in fact recovered also
in the strong coupling limit, we can study other examples of lattice
WI's, like for instance the axial WI for the $\langle ASP \rangle$
correlation function. In this case, current algebra requires: 
\renewcommand{\arraystretch}{2.0} 
\begin{equation}
\label{asp-wi} 
\begin{array}{l}
\nabla _\mu ^{Lx}\langle A_\mu ^k(x)S^j(y) P^l(z)\rangle = 2 \left( m- 
\widehat{m}\right)\, \dfrac{Z_mZ_P}{Z_A} \langle P^k(x)S^j(y)P^l(z)\rangle 
- \\ \qquad 
- d^{kjr}\,\delta _{x,y}\, \dfrac{Z_P}{Z_SZ_A} \langle P^r(x)P^l(z)\rangle
- d^{klr}\,\delta _{x,z}\, \dfrac{Z_S}{Z_PZ_A} \langle S^r(x)S^j(y)\rangle
\end{array}
\end{equation}
\renewcommand{\arraystretch}{1.5}An explicit calculation of the
correlation functions in the effective theory shows that again the WI
has the form prescribed by eq.~(\ref{asp-wi}). $Z_A$ is still given by
eq.~(\ref{za}) and $Z_m = 1/Z_P$. In addition, we also find the ratio: 
\begin{equation}
\label{zszp} \frac{Z_S}{Z_P}=1 
\end{equation}

As a last example, we wish to discuss a vector WI. As far as the vector 
current is concerned, we know that, for any value of the gauge coupling,
this current is conserved. Thus, its renormalization constant must be
exactly equal to 1. This can be verified by considering, for instance,
the vector WI for the $\langle VPP\rangle$ correlation function. In this
case, in the limit of degenerate quark masses, the prescription of
current algebra is:
\begin{equation}
\label{vpp-wi} 
\begin{array}{l}
Z_V \, \nabla _\mu ^{Lx}\langle V_\mu ^k(x)P^j(y) P^l(z)\rangle = 
   i\, f^{kjr}\,\delta _{x,y}\, \langle P^r(x)P^l(z)\rangle \ +
\\ \qquad \qquad \qquad \qquad \qquad \qquad
 + \ i\,f^{klr}\,\delta _{x,z}\, \langle P^r(x)P^j(y)\rangle
\end{array}
\end{equation}
This is exactly the identity we obtain in the effective theory, where,
as expected, we also find:
\begin{equation}
\label{zv} Z_V=1
\end{equation}

To conclude this section, we wish to notice that above results provide a
clear interpretation of eq.~(\ref{gell-mann2}). Since the combination
$Z_m Z_P$ is equal to 1, the right-hand side of eq.~(\ref{gell-mann2})
is renormalization invariant. In contrast, the pion decay constant
$F_{\pi}$ on the left-hand side of the equation must be renormalized,
the proper renormalization constant being $Z_A=1+r^2$. Thus, once
expressed in terms of all renormalized quantities, with $F_{\pi R} = Z_A
F_{\pi}=F_0$, eq.~(\ref{gell-mann2}) becomes:
\begin{equation}
\label{gell-mann3} F_{\pi R} ^2\,M_\pi ^2=2 m_R \langle \overline{\psi }
\psi \rangle _R
\end{equation}
that reproduces the well known continuum relation.

Thus, all the results obtained in this section are consistent with the 
following picture: at the leading order in the strong coupling and large
$N$ expansion, the lattice theory reproduces correctly the relations of
continuum current algebra. This allows one to define, even in the strong
coupling limit, renormalized lattice operators that corresponds to their
continuum counterpart. The lattice vector current, being conserved, does
not require renormalization. In contrast, the axial current does. The
resulting renormalized operators then respect all requirements of the
PCAC hypothesis. The way in which this scenario sets up is exactly the
one outlined in ref.~\cite{bochicchio} for the weak coupling regime of
lattice QCD.

\section{The ``renormalized'' effective Lagrangian\label{sec:rino}}

The discussion of the two previous sections has been mainly devoted to 
investigate the role of the Wilson terms (eq.~(\ref{SW})) in the lattice
effective Lagrangian. 

The original motivation, which led to the introduction of the Wilson
term in the lattice action, was the removing of the doubler fermion
species \cite{wilson}. In the naively discretized version of the Dirac
theory, these fermions would originate additional pseudo-Goldstone
bosons. Clearly, were these particles present in the spectrum of the
theory, they could not have been integrated out from the effective
Lagrangian. Thus in this sense, by having considered only a multiplet of
pseudo-Goldstone bosons, this primary effect of the Wilson term has been
taken into account, form the very beginning, in our lattice effective
theory.  

A well known consequence of removing the doubler species is the
appearance on the lattice of the axial anomaly \cite{karsten}. However,
in the effective theory considered in this paper, we have not introduced
external sources coupled to flavour-singlet anomalous axial
currents. Thus, the Wilson terms in our effective action do not play any
role in reproducing the correct anomaly. From this particular point of
view, they could have been completely neglected. 

Beside that, most of the other effects of the Wilson term on the lattice
are usually undesired, in that they contribute to further differentiate
the lattice theory from the corresponding continuum one. By being
formally vanishing in the continuum limit, the Wilson term in the
original lattice action can only induce physical effects through the
loops of the quantum theory. In the weak coupling region, these effects
can be cured by considering a proper set of renormalization
prescriptions, which allows one to relate the lattice quantities to the
corresponding continuum counterparts. Examples of such prescriptions,
already encountered in this paper, are the renormalization of the quark
mass and the multiplicative renormalization of the bilinear quark
operators. The several Wilson terms in the effective lattice Lagrangian
enter to exactly reproduce these ``undesired'' effects of the Wilson
term in the correlation functions of the lattice operators.

In this context, it appears natural to consider instead the following 
approach: on one hand, one can simply discard the Wilson terms in the
lattice effective theory, thus ending up with an effective Lagrangian
with a reduced number of terms each of those has a direct correspondence
in the continuum chiral Lagrangian. On the other hand, in comparing the
predictions of this effective theory with the results of a numerical
simulation, one must consider, in the latter, only the correlation
functions of properly renormalized operators.

In the following, we will show how this approach works in details by 
considering two specific examples. In doing that, we will take advantage
of the fact that, as discussed in the previous sections, from the
particular point of view of renormalization, the lattice theory at the
leading order in the strong coupling and large $N$ limit shows basically
the same features expected in the weak coupling region. Thus, we will be
able to continue our discussion by dealing with the effective lattice
Lagrangian derived in this approximation.

\subsubsection*{The quark mass renormalization}

In the limit of vanishing bare quark mass and Wilson term, the
generation of a quark mass from pure quantum corrections in the lattice
theory is protected by chiral symmetry. In contrast, in the presence of
the Wilson term, a physical quark mass can arise even in the limit in
which the mass term of the lattice action is set to zero. This is a well
known consequence of the Wilson breaking of chiral symmetry.

In this respect, the local external source $\chi (x)$ of the lattice
Lagrangian cannot be identified, directly, with the corresponding source
of the continuum theory. In fact, when in the latter this source is
removed, the pseudoscalar Goldstone bosons remain exactly massless. On
the lattice, instead, when we consider the limit $\chi (x)=0$,
corresponding to $m=-4r$, we still find a finite value of the pion mass
(precisely, from eq.~(\ref{coeffm}), $M_\pi^2 =-16r^2/(1+r^2)$). 

In order to define an external source that is proportional, in the
continuum limit, to the corresponding source of the continuum
Lagrangian, we can perform in the effective action the following change
of variable: 
\begin{equation}
\label{deltachi}  
\chi (x)=\left(\widehat{m}+4r\right) \xi (x)+\delta \chi (x)
\end{equation}
where $\widehat{m}$ is the quantity defined in eq.~(\ref{m-cap2}) and
the Wilson source $\xi (x)$ is introduced in eq.~(\ref{deltachi}) to
preserve chiral covariance. The new field $\delta\chi(x)$ is defined in
such a way that, in the limit of vanishing external sources, it reduces
to the renormalized quark mass:
\begin{equation}
\label{deltam}
\left( \delta \chi \right)_0 = \delta m=m-\widehat{m}
\end{equation}
For simplicity, we are only considering here the additive
renormalization of the quark mass, thus neglecting the existence of an
additional multiplicative renormalization.

By construction, after this change of variable the whole pion mass in
the effective theory is generated by the new source
$\delta\chi$. Precisely, as a consequence of the substitution
(\ref{deltachi}), a new term appears in the lattice effective action, of
the form:
\begin{equation}
\label{newterm}  
-\frac 12 N C_m \left(\widehat{m}+4r\right) \ \dsum\limits_x {\rm Tr} \left[ 
\xi ^{\dagger }(x)\,U(x)\,+\,\xi (x)\,U^{\dagger }(x)\right]
\end{equation}
The contribution to the pion mass coming from this term is exactly
canceled by an the opposite contribution coming from the Wilson
terms. As a result, the pion mass in the effective theory is only
generated by the term $S_m$ of eq.~(\ref{sum}), now containing the new
source $\delta\chi$. 

If in a specific case we were only interested in the calculation of the
pion mass, it is clear that, after the change of variables
(\ref{deltachi}), we could completely forget about all the Wilson terms
in the effective action. In addition it is worth to notice that, but for
the formal substitution $\chi (x) \rightarrow \delta\chi (x)$, the mass
term $S_m$ in the effective action is left unchanged. Thus in this
respect, we could have also considered, from the very beginning, a
completely different point of view. That is, we could have discarded all
the Wilson terms in the lattice effective action and simply interpreted
the source $\chi(x)$ as the ``renormalized" source associated with the
renormalized quark mass $\delta m$.

Technically, there is a difference between the two particular points of
view. This is due to the fact that, after the substitution
(\ref{deltachi}), the Wilson sources $\sigma(x)$ and $\pi(x)$ are
coupled in the action to different operators, namely $\Sigma
^{\prime}(x)$ and $\Pi ^{\prime}(x)$ given by: 
\begin{equation}
\label{newops}
\begin{array}{l}  
\Sigma ^{\prime} (x) = \Sigma (x) + \left(\widehat{m}+4r\right) \, S(x) \\
\Pi ^{\prime} (x) = \Pi (x) + \left(\widehat{m}+4r\right) \, P(x)  
\end{array}
\end{equation}
However, as far as we are not interested in calculating the correlation
functions of these particular operators, the above difference becomes 
completely immaterial.

Other interesting consequences of the change of variable defined by 
eq.~(\ref{deltachi}), and in particular those related to the lattice
chiral WI's, will be discussed at the end of this section. Before that,
we want to consider another significative example.

\subsubsection*{The axial current renormalization}

In the previous example we have shown that some of the Wilson terms in
the effective Lagrangian can be removed by considering an external
source which is directly coupled to a renormalized quantity, in that
specific case the quark mass. Similarly, we can argue that other Wilson
terms would be also removed by considering for instance a new axial
source, $a_\mu^{\prime}(x)$, directly coupled to the renormalized axial
current, $A_{R\mu} (x)=Z_A A_\mu(x)$. Specifically, we would like to
consider a change of variable leading to the introduction of the
following set of external sources:
\begin{equation}
\label{new_va}
\begin{array}{c}
W_\mu^{\prime} (x)=\exp \left\{ -i\left[ v_\mu^{\prime} (x)+a_\mu^{\prime} (x)
\right] \right\}  \\ 
Z_\mu^{\prime} (x)=\exp \left\{ -i\left[ v_\mu^{\prime} (x)-a_\mu^{\prime} (x)
\right] \right\} 
\end{array}
\end{equation}
Since the vector current is conserved, one might look for the new vector
and axial sources in the form:
\begin{equation}
\label{naive}
v_\mu(x) = v_\mu^{\prime}(x) \qquad, \qquad a_\mu(x) = Z_A a_\mu^{\prime}(x)   
\end{equation}
with $Z_A=$ given by eq.~(\ref{za}) in the strong coupling limit. 

This would imply a relation between the old and new $W_\mu$ and $Z_\mu$
external sources, which takes the form of an infinite series: 
\renewcommand{\arraystretch}{2.0}
\begin{equation}
\label{new_wz_naive} 
\begin{array}{l}
W_\mu (x)= W_\mu^{\prime}(x) + \dfrac{r^2}{2} \left[ W_\mu^{\prime}(x) -
Z_\mu^{\prime}(x) \right] + \ldots \\
Z_\mu (x)= Z_\mu^{\prime}(x) + \dfrac{r^2}{2} \left[ Z_\mu^{\prime}(x) -
W_\mu^{\prime}(x) \right] + \ldots
\end{array}
\end{equation}
\renewcommand{\arraystretch}{1.5}where the dots represent terms which
are at least quadratic in the external sources $v_\mu^{\prime}$ and
$a_\mu^{\prime}$. However, the changes of variables of
eqs.~(\ref{naive}) and (\ref{new_wz_naive}) cannot be implemented as
such, not being consistent with chiral covariance. Requiring the
validity of eq.~(\ref{naive}) is too restrictive, since adding to the
right-hand sides of that equation any term linear in the Wilson external
sources, $\sigma(x)$ and $\pi(x)$, would not change the desired
result. Indeed, even with this addition, the sources $v_\mu^{\prime}
(x)$ and $a_\mu^{\prime}(x)$ would still be coupled, in the lattice
action, to the currents $V_\mu(x)$ and $Z_A A_\mu(x)$
respectively. Similarly, we could also add any other term which is at 
least quadratic in any of the defined external sources. 

Taking advantage of this freedom one can obtain chirally covariant
relations between old and new sources, satisfying all the desired
requirements, by modifying eq.~(\ref{new_wz_naive}) in the following
manner:
\renewcommand{\arraystretch}{2.0}
\begin{equation}
\label{new_wz} 
\begin{array}{l}
W_\mu (x)= W_\mu^{\prime}(x) + \dfrac{r^2}{4} \left[ \xi(x) \xi^{\dagger}(x)
W_\mu^{\prime}(x) + W_\mu^{\prime}(x) \xi(x+\widehat\mu) \xi^{\dagger}(x+
\widehat\mu) - \right. \\ \qquad \qquad \left. 
- 2 \xi(x) Z_\mu^{\prime}(x) \xi^{\dagger}(x+\widehat\mu) \right]
+ \ldots \\
Z_\mu (x)= Z_\mu^{\prime}(x) + \dfrac{r^2}{4} \left[ \xi(x)^{\dagger} \xi(x)
Z_\mu^{\prime}(x) + Z_\mu^{\prime}(x) \xi^{\dagger}(x+\widehat\mu) \xi(x+
\widehat\mu) - \right. \\ \qquad \qquad \left. 
- 2 \xi^{\dagger}(x) W_\mu^{\prime}(x) \xi(x+\widehat\mu) \right]
+ \ldots
\end{array}
\end{equation}
\renewcommand{\arraystretch}{1.5}In this expressions we have included on
the right-hand side all the terms which are at most linear in the
$v_\mu^{\prime}$, $a_\mu^{\prime}$, $\sigma$ and $\pi$ external sources,
whereas the dots represent an infinite series of terms at least
quadratic in these sources. The missing terms can be determined, order
by order, by requiring the unitarity of the new fields
$W_\mu^{\prime}(x)$ and $Z_\mu^{\prime}(x)$. By expanding
eq.~(\ref{new_wz}) in terms of vector, axial, scalar and pseudoscalar
sources, we find, up to higher orders:
\begin{equation}
\label{not_naive}
\begin{array}{l}
v_\mu(x) = v_\mu^{\prime}(x) + O(j^2) \\
a_\mu(x) = \left(1+r^2\right) a_\mu^{\prime}(x) - \dfrac{r^2}{2} \nabla
_\mu \pi (x) + O(j^2)
\end{array}
\end{equation}
which generalize eq.~(\ref{naive}).

It is worth to discuss an interesting consequence of the combined change
of variables (\ref{deltachi}) and (\ref{not_naive}). Once both these
changes are performed, the terms in the effective action which are
linear in the external sources can be still written in the form
(\ref{sphi1}) and (\ref{effop}), but now with new coefficients $d_i$
given by: 
\begin{equation}
\label{coeff1_new} 
\begin{array}{l}
d_s=-n N C_m \quad ,\quad d_{s1}=NC_m/F_0^2 \\
d_p=-NC_m/F_0 \quad ,\quad d_a=-F_0 \quad ,\quad d_v=1 \\ 
d_\sigma =8nr^2 \left[NC_1 - F_0^2/\left(1+r^2\right)\right] \quad ,\quad 
d_{\sigma 1} = 0 \\
d_{\pi 1}=0 \quad ,\quad d_{\pi 2}=0 \\
\end{array}
\end{equation}
In particular, because of the last two equalities ($d_{\pi 1}=d_{\pi 2}=0$), 
the operator:
\begin{equation}
\label{piprime}
\Pi ^{\prime} (x) = \Pi (x) + \left(\widehat{m}+4r\right) \, P(x) +
\frac{r^2}{2} \nabla _\mu ^L A_\mu (x)
\end{equation}
that is now coupled to the source $\pi(x)$ in the lattice action, has
always vanishing on-shell matrix elements, and can only contribute to
the lattice correlation functions through localized contact terms. In
some cases it does not contribute at all. For instance, the lattice
chiral WI for the $\langle AP \rangle$ correlation function can be now
directly derived in the form (\ref{ap-wi4}), without any explicit
additional contribution from the Wilson terms.

\subsubsection*{The continuum-like effective Lagrangian}

The above discussion naturally leads to the following conclusion: since
it is possible, both in the weak and strong coupling limit, to
renormalize the lattice correlation functions in such a way that they
satisfy all the relations of continuum current algebra, it must be also
possible to describe the physics of this renormalized lattice theory in
terms of a ``renormalized'', continuum-like, effective
Lagrangian. Specifically, the only sources of chiral symmetry breaking
in this Lagrangian must be represented by the quark mass terms, whereas
Wilson terms like those in eq.~(\ref{SW}) should not enter at all.

The two examples discussed in this section were aimed to support the
above argument. An evident feature of eq.~(\ref{coeff1_new}) is that,
after the external sources renormalization, many of the coefficients
related to the Wilson operators $\Sigma(x)$ and $\Pi(x)$ do actually
vanish. In addition, the Wilson parameter $r$ does not explicitly enter
anymore in the expressions of the remaining coefficients, related to the
lattice operators $V_\mu$, $Z_A A_\mu$, $S$ and $P$. In fact, one can
easily show that after the change of variables (\ref{deltachi}) and
(\ref{new_wz}) many of the correlation functions we have considered so
far are simply reproduced by the following effective action:
\begin{equation}
\label{SX}
S(U)=\dsum\limits_{x}{\rm Tr }\left[ \dfrac 14 F_0^2 \, \left(\Delta _\mu U
\right) ^{\dagger } \left(\Delta _\mu U\right) - \dfrac 12 N C_m \left(\chi 
^{\dagger}U+ U^{\dagger } \chi \right) \right]
\end{equation}
provided the external sources $W_\mu(x)$, $Z_\mu(x)$ and $\chi(x)$
entering in it are identified with the renormalized sources of
eqs.~(\ref{deltachi}) and  (\ref{new_va}). Equation (\ref{SX}) exactly
corresponds, up to discretization effects, to the continuum chiral
Lagrangian at order $p^2$ \cite{gl}. 

On the other hand, it should be pointed out that a continuum-like
effective action, like that of eq.~(\ref{SX}), cannot be obtained from
the original one by simply renormalizing the external sources. For
instance, the correlation functions of the operator $A_\mu^2(x)$
obtained from eq.~(\ref{SX}) are not equal to the corresponding
correlation functions given by the lattice effective Lagrangian after
the renormalization of the axial source. The reason is that the lattice
the operator $A_\mu^2(x)$ does not renormalize with renormalization
constant $Z_A^2$, and in fact it does not even renormalize in a simple
multiplicative way. Only some of the effects of the Wilson terms in the
lattice effective theory can be accounted for by a renormalization of
the external sources, because these effects rather imply the
renormalization of an infinite number of lattice operators. The
important point here is the following: with a proper renormalization of
the lattice operators, their correlation functions will be correctly
reproduced, up to pure discretization effects, by a continuum-like
effective lattice Lagrangian which does not contain the Wilson terms at
all. The reason is that these correlation functions satisfy all the WI's
predicted by continuum current algebra, and these identities can be only
reproduced by a continuum-like effective Lagrangian. 

\section{A numerical calculation of the lattice effective Lagrangian in
the scaling region\label{sec:scaling}}

In the region of intermediate couplings, which is relevant for numerical
simulations of continuum QCD, the non-effective degrees of freedom of
the lattice theory cannot be analytically integrated out. Therefore, in
this case we would not be able to derive the exact form of the resulting
effective theory. However, such an integration can be performed
numerically, at least in an approximate way. 

The idea consists in defining an effective lattice Lagrangian of the
general form derived in this paper for the strong coupling and large $N$
limit. The basic fields entering in this Lagrangian are the effective
field $U(x)$ and the set of external sources $W_\mu(x)$, $Z_\mu(x)$ and
$\chi(x)$. On the basis of the results derived in this paper, we expect
that the correlation functions of properly renormalized lattice
operators can be reproduced by an effective lattice Lagrangian which
does not necessitate Wilson terms. Furthermore, the external sources of
this effective Lagrangian have a direct and well defined correspondence
with the external sources entering in the continuum QCD chiral
Lagrangian.

The most general form of the effective lattice Lagrangian is dictated by
local chiral invariance, lattice symmetries, parity and charge
conjugation. The chiral parallel transport between two nearest neighbor
lattice sites is performed by the left- and right-handed external
sources $Z_\mu(x)$ and $W_\mu(x)$ respectively. They can enter in the
Lagrangian in the form of chiral plaquettes or through the lattice
covariant derivative $\Delta _\mu U(x)$ defined in
eq.~(\ref{deltamu}). The local source $\chi(x)$, which reverses
chirality, allows to introduce in the theory the effects of chiral
symmetry breaking induced by the light quark mass term.

The general strategy to calculate the effective lattice Lagrangian has
been outlined in ref.~\cite{cs}. The idea is to assume a sufficiently
large set of effective couplings, with strength determined by unknown
numerical coefficients. These coefficients can be then fixed through the
matching of an overcomplete set of expectation values, computed both in
the effective and the full theory. 

The calculation in the full theory, being highly non perturbative,
requires the implementation of a numerical lattice simulation. An open
question is whether or not the lattice effective Lagrangian, in the
region of intermediate values of the lattice QCD coupling constant, can
be treated in perturbation theory. This is possible in the strong
coupling and large $N$ limit, where the coefficients of the higher order
terms ($\phi^4$, $\phi^6$, $\ldots$) are suppressed by increasing powers
of $1/N$. In addition, even for the actual value $N=3$, these
coefficients (see e.g. eq.~(\ref{coeff0})) are found to lie in the
perturbative regime. Certainly, such a feature could be lost as one
approaches the weak coupling region. Even if this should be the case,
however, the difficulty could be more technical than conceptual, in the
sense that also the effective theory could be treated by numerical
techniques. 

A crucial observation is that, since the lattice effective Lagrangian
contains explicitly the collective fields responsible for the long
distance behaviour of the fundamental lattice theory, then only short
distance couplings are expected to play a relevant role in the effective
theory. Therefore, in defining a criterion to truncate the infinite
number of allowed couplings in the effective Lagrangian one can limit
oneself to considering local or quasi-local interactions (e.g. only
local, nearest and next-to-nearest neighbors couplings). For the same
reason, the numerical simulation in the fundamental theory, which is
necessary to perform the matching, should be feasible on lattices of
moderate sizes, which would allow one to achieve a higher numerical
accuracy. 

In the strong coupling limit, the coefficients $C_k$ of the effective 
Lagrangian are found to be explicit functions of the light quark mass
$m$ and the Wilson parameter $r$. Both these parameters enter in the
coefficients through the vacuum expectation value $u_0$. Such a
dependence persists even after the process of renormalization, but there
are reasons to believe that it is an artifact of the strong coupling
expansion. In the weak coupling limit, the matrix elements of the
lattice renormalized operators must be independent on the value of
$r$. Therefore, the coefficients of the renormalized effective
Lagrangian cannot depend on the Wilson parameter. Furthermore, there is
clear phenomenological evidence that the effects of the light quark
masses, in the physical amplitudes, are well reproduced in the chiral
QCD Lagrangian by the terms containing the external source $\chi(x)$,
without any additional mass dependence entering in the coefficients. For
these reasons, in defining an effective lattice Lagrangian for the weak
coupling region one should assume the coefficients independent on $m$
and $r$. 

Once one has been able to determine an expression for the lattice
effective Lagrangian which accurately reproduces the Green's functions
of the fundamental lattice theory at moderate distances, one can compute
its large volume limit and expand in increasing powers of the external
momenta. The result of such an expansion, to any given order in $p$,
corresponds to the continuum low-energy QCD effective theory. In such a
way, the procedure discussed in this paper would allow a first principle
theoretical calculation of the coefficients of the QCD chiral Lagrangian.

\section*{Acknowledgments}

This research was supported in part under DOE grant DE-FG02-91ER40676.
V.L. acknowledges the support of an INFN post-doctoral fellowship.


\begin{thebibliography}{9}

\bibitem{wein1} S.Weinberg, Phys.Rev.Lett. {\bf 18} (1967) 188;
Phys.Rev. {\bf 166} (1968) 1568.   

\bibitem{wein2} S.Weinberg, Physica {\bf 96A} (1979) 327.  

\bibitem{gl} J.Gasser and H.Leutwyler, Annals of Phys. {\bf 158} (1984)
142; Nucl.Phys. {\bf B 250} (1985) 465.

\bibitem{meiss} U.G.Mei{\ss}ner, Rep.Prog.Phys. {\bf 56} (1993) 903,
hep-ph/9302247.  

\bibitem{pich} A.Pich, Rep.Prog.Phys. {\bf 58} (1995) 563,
hep-ph/9502366.  

\bibitem{ecker} G.Ecker, Prog.Part.Nucl.Phys. {\bf 35} (1995) 1, 
hep-ph/9501357. 

\bibitem{beg} J.Bijnens, G.Ecker and J.Gasser, The DA$\Phi$NE Physics 
Handbook (2nd edition), eds. L.Maiani et al., Frascati (1994),
hep-ph/9411232.

\bibitem{gm} H.Georgi and A.Manohar, Nucl.Phys. {\bf B 234} (1984) 189. 

\bibitem{res} G.Ecker et al., Nucl.Phys. {\bf B 321} (1989) 311.

\bibitem{cs} S.Myint and C.Rebbi, Nucl.Phys. {\bf B 421} (1994) 241,
hep-lat/9401009.

\bibitem{klu} H.Kluberg-Stern et al., Nucl.Phys. {\bf B 190} (1981) 504.

\bibitem{ks} N.Kawamoto and J.Smit, Nucl.Phys. {\bf B 192} (1981) 100.

\bibitem{wilson} K.Wilson, in {\it ``New phenomena in subnuclear
physics''}, ed. A.Zichichi (Plenum, New York, 1977).

\bibitem{bochicchio} M.Bochicchio et al., Nucl.Phys. {\bf B 262} (1985)
331. 

\bibitem{ape-ukqcd}  L.Lellouch et al., the UKQCD Collaboration,
Nucl.Phys. {\bf B 444} (1995) 401, hep-lat/9410012; C.R.Allton et al.,
the APE Collaboration, Phys.Lett. {\bf B 345} (1995) 513,
hep-lat/9411011.

\bibitem{bkt} N.Cabibbo, G.Martinelli and R.Petronzio, Nucl.Phys. {\bf B
244} (1984) 381; \\ G.Martinelli, Phys.Lett. {\bf B 141} (1984) 395.

\bibitem{bkn} M.Crisafulli et al., Phys.Lett. {\bf B 369} (1996) 325,
hep-lat/9509029. 

\bibitem{karsten}  L.H.Karsten and J.Smit, Nucl.Phys. {\bf B 183} (1981) 103.
  
\end{thebibliography}
\end{document}